\documentclass[12pt,preprint]{aastex}
\usepackage{graphicx}
\usepackage{subfigure}
\usepackage{subfloat}
\usepackage{longtable}
\usepackage{multirow}
\usepackage{amsmath}
\usepackage{color}
\usepackage{url}
\usepackage{lipsum}
\usepackage{booktabs}
\newcommand{\herschel}{{\it{Herschel}}}

\newcommand{\mum}{$\mu$m}

\begin{document}
\title{The Physical Characteristics of Interstellar Medium in NGC 3665 with \emph{Herschel} Observations$^\star$}
\author{Meng-Yuan Xiao\altaffilmark{1,2,5}, Yinghe Zhao\altaffilmark{3,4,6}, Qiu-Sheng Gu\altaffilmark{1,2,5}, Yong Shi\altaffilmark{1,2,5}}

\affil{$^1$ School of Astronomy and Space Science, Nanjing University, Nanjing 210093, P.~R.~China; \email{qsgu@nju.edu.cn}}
\affil{$^2$ Key Laboratory of Modern Astronomy and Astrophysics, Nanjing University, Nanjing 210093, P.~R.~China}
\affil{$^3$ Yunnan Observatories, CAS, Kunming 650011, P.~R.~China}
\affil{$^4$ Key Laboratory for the Structure and Evolution of Celestial Objects, CAS, Kunming 650011, P.~R.~China}
\affil{$^5$ Collaborative Innovation Center of Modern Astronomy and Space Exploration, Nanjing 210093, P.~R.~China}
\affil{$^6$ Center for Astronomical Mega-Science, CAS, 20A Datun Road, Chaoyang District, Beijing 100012, P.~R.~China}
\altaffiltext{$\star$}{\herschel\ is an ESA space observatory with science instruments provided by European-led Principal Investigator consortia and with important participation from NASA.}

\begin{abstract}
We present the analysis of the physical properties of the interstellar medium (ISM) in the nearby early-type galaxy NGC 3665, based on the far-infrared (FIR) photometric and spectroscopic data as observed by the Herschel Space Observatory. The fit to the spectral energy distribution reveals a high dust content in the galaxy, with the dust-to-stellar mass ratio of $M_{\rm dust}$/$M_\ast$ $\sim$ 1.1 $\times$ 10$^{-4}$ that is nearly three times larger than the mean value of local S0+S0a galaxies. For the ionized regions (H~\textsc{ii} regions), the electron density (n$_e$) is around 49.5 $\pm$ 11.9 cm$^{-3}$ based on the [N~\textsc{ii}]\,122\,\mum/[N~\textsc{ii}]\,205\,\mum\ ratio. For the photodissociation regions, the heating efficiency is in the range of 1.26 $\times$ 10$^{-3}$ and 1.37 $\times$ 10$^{-3}$ based on the ([C~\textsc{ii}$]+[$O~\textsc{i}]\,63\,\mum)/$L_{\rm TIR}$, which is slightly lower than other local galaxies; the hydrogen nucleus density and the strength of FUV radiation field are n $\sim$ 10$^{4}$ cm$^{-3}$ and G$_0$ $\sim$ 10$^{-0.25}$, respectively. The above results are consistent with the presence of weak AGN and a low level of star-forming activity in NGC 3665.  Our results give strong support to the `morphological quenching' scenario, where a compact, massive bulge can stabilize amount of cool gas against star formation.

\end{abstract}


\section{Introduction}

Interstellar medium (ISM) plays a crucial role in the galaxy formation and evolution. It is the primary reservoir for star formation. The mutual interaction between ISM and stars determines the rates of both gaseous depletion and star formation in the galaxy.
ISM contains several main components: ionized gas, neutral gas, cold molecular clouds, as well as dust grains. 
When molecular clouds collapse under their own gravity to form stars,  the gravity is required to overcome the random motion pressure, which requires the interstellar gas to be cooled down sufficiently. Neutral and ionized gas can be heated up by the photoelectric process \citep[PE;][]{Tielens1985}, and cooled via collisional excitation of C$^+$, O, N$^+$ and other elements. The atomic fine-structure emission lines in the far-infrared (FIR), such as [N~\textsc{ii}] 122 and 205 $\mu$m, [C~\textsc{ii}] 158 $\mu$m, [O~\textsc{i}] 63 $\mu$m, and [C~\textsc{i}] 370 $\mu$m, are very important coolants, which play a crucial role in the thermal balance in the H II regions and photodissociation regions (PDRs), and can be served as critical diagnostic tools for studying the physical properties of ISM \citep[e.g.,][]{Kaufman1999}.

Among these fine-structure lines, [C~\textsc{ii}] 158 $\mu$m is the brightest and the dominant FIR cooling line, typically accounting for 0.1\%-1\% of the total FIR luminosity \citep{Stacey1991, Malhotra2001, Diaz-Santos2013, Sargsyan2014}. The ionization potential of carbon is 11.26 eV, thus the [C~\textsc{ii}] line can be a tracer of both ionized and neutral gas. The [O~\textsc{i}] emission originates in the PDRs since the ionization potential of oxygen is just above 13.6 eV, whereas the [N~\textsc{ii}] line exclusively arises from the ionized gas as nitrogen has an ionization potential of 14.5 eV. As shown in previous studies \citep{Zhao2013, Zhao2016a, Sargsyan2014}, both [N~\textsc{ii}] 205 $\mu$m and [C~\textsc{ii}] 158 $\mu$m lines can be served as useful indicators of star formation rate (SFR). Based on the critical densities of collisional excitations for [N~\textsc{ii}] 122 and 205 $\mu$m, $\sim$293 cm$^{-3}$ and 44 cm$^{-3}$ at the electron temperature of 8000 K, respectively, the [N~\textsc{ii}] 122/[N~\textsc{ii}]  205 ratio is a sensitive probe of the electron density (10 $\lesssim$ $n_e$ $\lesssim$ 300 cm$^{-3}$) of ionized gas, which can be further used to estimate the fraction of [C~\textsc{ii}] 158 $\mu$m originating from ionized gas \citep{Oberst2006,Oberst2011}.

Based on \emph{Infrared Space Observatory} (\emph{ISO}) observations, \citet{Malhotra2000} investigated the physical properties of gas and dust in four elliptical/S0 galaxies, and proposed that softer radiation field might result in lower [C~\textsc{ii}]/$F_{\rm FIR}$ ratios than those of normal star-forming galaxies by a factor of $2 - 5$. With data of \emph{Herschel Space Observatory} \citep[hereafter {\it Herschel};][]{pil10}, \citet{Lapham2017} presented spectroscopic observations of the FIR emission lines in 20 nearby elliptical/S0 galaxies and found that the average of [C~\textsc{ii}]/$F_{\rm FIR}$ ratio is slightly lower than that of spiral galaxies, and [C~\textsc{ii}] luminosity can be served as a good SFR tracer for both early-type galaxies (ETGs) and spirals. Furthermore, \citet{Lapham2017} showed that the fraction of [C II] emission arising from ionized gas is similar in ETGs (63.5 \%) and normal spirals (53.0 \%). 

In this work, we investigate the ISM properties of a nearby ETG, NGC 3665. 
In general, ETGs in the local Universe contain very little cool gas and/or dust. Many studies focused on the mechanisms of star formation suppression for nearby ETGs, such as removing cool gas \citep{Di2005, Hopkins2006}, suppressing gas infall and cooling \citep{Birnboim2003, Croton2006, Lanz2016}, or stabilizing gas reservoirs \citep{Martig2009, Tacchella2015}, etc.
However, several recent works have shown that $\sim40\%$ elliptical and S0 galaxies are rich in both dust and cool gas \citep{Temi2007a, Temi2007b, Young2011, Young2014, Alatalo2013}, with the SFR reaching a few $M_\odot\,{\rm yr}^{-1}$ \citep{Temi2009a, Temi2009b}.

NGC 3665 (11$^{\mathrm{h}}$24$^{\mathrm{m}}$43$^{\mathrm{s}}$.7, $+38^{\circ}45\arcmin46\arcsec$) is located 33.1 Mpc away \citep{Cappellari2011}, with an inclination angle, $i$, of 69.9$^{\circ}$ \citep{Onishi2017}, and Hubble type of SA0 \citep[RC3;][]{deVauc91}. It is found to be gas-rich in the center \citep{Young2011,Serra2012, Alatalo2013, Davis2013, Davis2014, Kamenetzky2016, Nyland2016, Onishi2017}, with a total molecular gas mass of $10^{9.11\pm0.01}$ $M_\odot$ from the Combined Array for Research in Millimeter Astronomy (CARMA) observations \citep{Alatalo2013}. \citet{Alatalo2013} showed a prominent dusty disk in the $g-r$ image, and a regularly rotating central molecular gas disk in the CO ($1-0$) emission. \citet{Onishi2017} obtained similar results of a prominent dust structure in the \emph{Hubble Space Telescope} (HST) $H$-band image, and a centrally-concentrated gaseous disk in the CO($2-1$) observations, and derived the mass of the central super-massive black hole of 5.75 $\times$ 10$^8$ $M_\odot$. In \citet{Davis2014}, NGC 3665 shows the largest deviation from the so-called Kennicutt-Schmidt relation \citep{Kennicutt1998a} among 32 CO-detected ATLAS$^{\rm 3D}$ ETGs. The abundant cool gas and low SFR surface density ($\Sigma_{\rm SFR}$) suggest that NGC 3665 is an atypical ETG.  Through studying specific ISM properties, we can better understand its relation to the star formation, especially the reasons for such a low star formation efficiency.

In this paper, we focus on the photometric observations of NGC 3665 at 100, 160, 250, 350 and 500 $\mu$m, obtained with the Photodetector Array Camera and Spectrometer \citep[PACS;][]{Poglitsch2010} and the Spectral and Photometric Imaging REceiver \citep[SPIRE;][]{Griffin2010} onboard \emph{Herschel}, with five fine-structure lines of [N~\textsc{ii}] 122 and 205 $\mu$m, [C~\textsc{ii}] 158 $\mu$m, [O~\textsc{i}] 63 $\mu$m, and [C~\textsc{i}] 370 $\mu$m. We also perform optical spectroscopic observations of NGC 3665 with the CAHA 3.5m telescope. Using these spectra combined with the analysis of multi-wavelength spectral energy distribution (SED), we investigate SFR, the gas heating and cooling efficiency in PDRs, and the hydrogen nucleus density and the strength of FUV radiation field derived from the PDR models. 

This paper is organized as follows. Data reductions for \emph{Herschel} photometric and spectroscopic observations and CAHA 3.5m telescope spectroscopic observations are described in Section 2. We present the results and discuss their implications in Section 3, including the classification of nuclear activity, SED fitting, star formation rate, the ionized gas contribution to the [C~\textsc{ii}] 158 $\mu$m emission, the photoelectric heating efficiency of the interstellar gas, and some derived values using the PDR model. Then, we compare our results with respect to the star formation and gas to the previous works, and discuss the possible mechanisms to suppress star formation in NGC 3665 in Section 4. The conclusions for this work are summarized in Section 5.

\section[]{Observations and Data Reduction}
\subsection{\emph{Herschel} PACS and SPIRE Photometry}

NGC 3665 was observed with \emph{Herschel} PACS and SPIRE in two open time projects GT2\_mbaes\_2 (PI: M. Baes) and OT1\_lyoung\_1 (PI: L. Young), respectively.
We performed aperture photometry with the public Level 2 map products, which were downloaded from the Herschel Science Archive\footnote{\url{http://herschel.esac.esa.int/Science_Archive.shtml}}.
To match the $36''$ resolution of the 500 $\mu$m image, we convolved other band images with the kernels provided in \citet{Gordon2008}. We adopted the aperture of $40''$ radius around the source, and determined the sky values within 60-$90''$ annulus surrounding the target galaxy. 
The photometric errors on the source were given by the standard deviation of 15 annulus between $60''$ and $90''$ around the source, as well as the 5\% uncertainty of the fiducial stellar models in the PACS photometry and the confusion noise in the SPIRE photometry (Herschel Observers' Manual\footnote{\url{http://herschel.esac.esa.int/Docs/Herschel/html/Observatory.html}}, v.5.0.3, 2014). We then applied the corresponding color and aperture corrections to these fluxes. The final photometric fluxes are given in Table 1. 

\subsection{\emph{Herschel}  PACS and SPIRE Spectroscopy}

The \emph{Herschel} FIR spectroscopic observations of NGC 3665 were performed by the program OT1\_lyoung\_1 \citep[PI: L. Young;][]{Lapham2017}, with a total exposure time of 60.7 hours  of {\it Herschel}.
We focused on the [N~\textsc{ii}] 122 and 205 $\mu$m, [C~\textsc{i}] 370 $\mu$m, [C~\textsc{ii}] 158 $\mu$m and [O~\textsc{i}] 63 $\mu$m fine-structure lines observed with the PACS (better than $12''$) and SPIRE ($\sim17''$ and $\sim36''$) instruments. 

\subsubsection{PACS flux measurement}

The PACS integral-field spectrometer covers the 51-220 $\mu$m range with a spectral resolution of $\sim$75-300 km s$^{-1}$, which is composed of $5\times5$ squared spatial pixels (spaxels), each with a size of 9$^{\prime\prime}.4$. It covers a total projected field of view (FoV) of $47''\times47''$. The angular resolutions are 9$^{\prime\prime}.4$ at $\sim$63 $\mu$m, 10$^{\prime\prime}.0$ at $\sim$122 $\mu$m, and 11$^{\prime\prime}.5$ at $\sim$158 $\mu$m. For NGC 3665, the [O~\textsc{i}] 63 $\mu$m, [N~\textsc{ii}] 122 $\mu$m and [C~\textsc{ii}] 158 $\mu$m data are comprised of observations acquired in the mapping mode.
The Level 2 data with standard rebinned cubes, observed with line range spectroscopy mode and reduced with the Herschel Interactive Processing Environment \citep[HIPE;][]{Ott2010} version 14.2, were downloaded directly from the Herschel Science Archive.  The spectrometer effective spectral resolutions are about 86 km s$^{-1}$ (the third grating order), 290 km s$^{-1}$ (the first grating order) and 238 km s$^{-1}$ (the first grating order) for [O~\textsc{i}] 63 $\mu$m, [N~\textsc{ii}] 122 $\mu$m and [C~\textsc{ii}] 158 $\mu$m, respectively. The basic observational informations for each line are summarized in Table 2.

Following \citet{Farrah2013}, we co-added the spectra of the central 9 spaxels to measure the line fluxes, due to the fact that the line-emitting regions in NGC 3665 are not confined in the central spaxel according to its 100 micron continuum emission.
Then we fitted the observed spectrum with two Gaussian functions (for the line) plus a linear component (for the continuum), as shown in Fig. 1. The uncertainty of the integrated flux was calculated according to the rms of the continuum. A point-source aperture correction \citep{Balog2014} was applied to derive the final flux. The intrinsic line width ($\sigma_{\rm true}$) was obtained using $\sigma_{\rm true} = \sqrt{\sigma_{\rm obs}^2 - \sigma_{\rm inst}^2}$, where $\sigma_{\rm obs}$ and $\sigma_{\rm inst}$ are the observed line width and instrumental spectral resolution, respectively. 
The [O~\textsc{i}] 63 $\mu$m is not detected, and thus we calculated its upper limit using 3$\sigma$ times the instrumental spectral resolution at the considered wavelength. The line fluxes are given in Table 2.
In Fig. 2, we show the [C~\textsc{ii}] 158 $\mu$m and [N~\textsc{ii}] 122 $\mu$m integrated intensity maps, overlaid with the $^{12}$CO$(J=1-0)$ moment 0 contours from the CARMA ATLAS$^{\rm 3D}$ survey \citep{Alatalo2013}. The maps are created by projecting the rasters onto a common, regular spatial grid with $1.86''$ and $1.63''$ pixel sizes, respectively, clearly shows that the neutral and ionized gas has extended structures and follow the CO($1-0$) gas disk. Both of these two cooling lines are strongest at the center, and weaker outwards, suggesting a mount of [C~\textsc{ii}] 158 $\mu$m lines generate from ionized gas, which emit the [N~\textsc{ii}] 122 $\mu$m. 
The offset on the nucleus in each panel of [C~\textsc{ii}] 158 $\mu$m and [N~\textsc{ii}] 122 $\mu$m lines compared with CO($1-0$) contour is smaller than $2''$ and within the beam size. Since the non-detection of [O~\textsc{i}] 63 $\mu$m, and the [N~\textsc{ii}] 122 $\mu$m emission is marginally resolved, we tried but failed to perform radial decomposition. Thus, we only concentrated on the analysis of the integrated properties of NGC 3665.

\subsubsection{SPIRE flux measurement}            
The SPIRE Fourier-Transform Spectrometer (FTS) consists of two bolometer detector arrays, the SPIRE Short Wavelength Spectrometer Array (SSW) and the SPIRE Long Wavelength Spectrometer Array (SLW), covering overlapping bands of 191$-$318 $\mu$m and 294$-$671 $\mu$m, respectively. The observations were conducted in the single pointing mode with a high spectral resolution of 0.04 cm$^{-1}$ (or 1.2 GHz in frequency space). We used the Level 2 data products, which were reduced using the standard pipeline provided by the HIPE version 14.0, along with the SPIRE calibration version 14.2. 

To obtain the integrated fluxes of [C\,\textsc{i}] and [N~\textsc{ii}] 205 $\mu$m lines, we followed the method in \citet{Lu2017}. 
For the [C~\textsc{i}] 370 $\mu$m line, we adopted a pure \emph{Sinc} function as (1) this line is only partially resolved given the large instrumental resolution ($\sim$540 km s$^{-1}$) at 370 $\mu$m, and (2) the S/N is not high enough to use a Sinc-convolved-Gauss function. For the [N~\textsc{ii}] 205 $\mu$m line, we adopted a Sinc-convolved-Gauss function as the velocity resolution at $\sim$205 $\mu$m is about 300 km s$^{-1}$. We also used \emph{FTLinefitter}\footnote{https://www.uleth.ca/phy/naylor/index.php?page=ftfitter} (FTFitter; in version 1.9) to fit the observed spectra, and obtained similar results. Table 2 gives the line fluxes of [N~\textsc{ii}] 205$\mu$m and [C~\textsc{i}] 370 $\mu$m.

The SPIRE beam size at 205 $\mu$m is $\sim 17''$, and thus might not cover the total emission of the [N~\textsc{ii}] line.  To check this, we used the correlation between the [N~\textsc{ii}] 205 emission to total infrared luminosity ratio ($L_{\rm [N\,{\scriptsize \textsc{ii}}]205\,\mu{\rm m}}/L_{\rm TIR}$; see Section 3.2 for the calculation of $L_{\rm TIR}$ used here) and FIR color \citep{Zhao2016a}: $\log (L_{\rm [N\,{\scriptsize \textsc{ii}}]205\,\mu{\rm m}}/L_{\rm TIR}) = -3.83$ $-$ 1.26$x$ $-$ 1.86$x^{2}$ $-$ 0.90$x^{3}$, where $x = \log (f_{\rm 70}/f_{\rm 160})$,  and $f_{\rm 70}$ and $f_{\rm 160}$ represent the flux densities at 70 and 160 $\mu$m, respectively. We obtained the total [N~\textsc{ii}] 205 emission is $\sim$ 6.12 $\times$ 10$^{-17}$ W m$^{-2}$, consistent with the measured value within the uncertainties. 
Furthermore, the [N~\textsc{ii}] flux is 7.29 $\pm$ 0.14 $\times$ 10$^{-17}$ W m$^{-2}$ after calibrating with the Semi Extended Correction Tool (SECT) in \citet{Lapham2017}. Therefore, the distribution of [N~\textsc{ii}] 205 $\mu$m emission should not deviate much from a point-like source relative to the 17\arcsec\ beam.

\subsection{Optical Spectroscopy}
The optical spectroscopic observations were carried out with the 3.5-m telescope at Calar Alto Observatory (CAHA\footnote{http://www.caha.es/}, Almer\'ia, Spain) on May 26, 2017. We used the PPAK integral-field unit (IFU) \citep{Verheijen2004} at Potsdam Multi-Aperture Spectrometer \citep[PMAS;][]{Roth2005} instrument, containing 382 fibers with each of $2.7''$ diameter and a hexagonal FoV of $74''$ $\times$ $64''$ \citep{Kelz2006}. The V500 grating, which has a spectral resolution (FWHM) of 6 {\AA} and wavelength coverage of 3745-7500 {\AA}, was adopted. A three-pointing dithering scheme was used with exposure time of 900 seconds each. The typical airmass was $\sim$1.1.  
Raw spectroscopic data was reduced with the RGB reduction pipeline version 0.0.2, including bias subtraction, flat-field correction, cosmic ray removal, atmospheric extinction correction, as well as wavelength and flux calibration. The wavelength calibration was performed based on Hg/He lamp exposures at the beginning of the observations every time.  For the following analysis, we only focus on the central $3''$ $\times$ $3''$ (9 spaxels).

To obtain the emission line fluxes, we followed the method of \citet{Tremonti2004} and \citet{Brinchmann2004} to model the stellar continuum with templates, which are generated using the popular synthesis code of \citet[BC03]{BC2003} . The template spectra are composed of ten different ages (0.005, 0.025, 0.1, 0.2, 0.6, 0.9, 1.4, 2.5, 5, 10 Gyr) and four metallicities (0.004, 0.008, 0.017, and 0.05). For each metallicity, we performed a non-negative least square fit to obtain the best-fitting model spectrum using the 10 single-age populations, with the internal dust attenuation model of \citet{Charlot2000}. During the fitting process, each template is convolved with a stellar velocity dispersion from 0 to 200 km s$^{-1}$ by a step size of 5 km s$^{-1}$.  After subtracting the best-fitting stellar continuum model, we obtained the pure nebular emission line spectrum, and fitted each line with one Gaussian component, including H$\beta$, H$\alpha$, [{O~\sc iii}]$\lambda$5007, and [{N~\sc ii}]$\lambda$6584 lines.

\section[]{Results and Discussion}
\subsection{Spectral Classification}
To identify the power source of the emission lines, we adopted the well-known BPT diagnostic diagram \citep{Baldwin1981, Veilleux1987}. Here we only focus on the central $3''$ $\times$ $3''$ region (the IFU data will be fully used in the following paper for a large sample of S0 galaxies) in NGC 3665. In Fig. 3, we plotted [{N~\sc ii}]$\lambda$6584 /H$\alpha$ versus [{O~\sc iii}]$\lambda$5007/H$\beta$ flux ratios with ${\rm S/N} >3$ for all of the four emission lines. 
The red solid and dashed lines mark the criteria to separate AGN from star-forming galaxies according to \citet{Kewley2001} and \citet{Kauffmann2003}, respectively, with composite systems located in between these two lines. The horizontal blue line adopted from \citet{Kauffmann2003} is used to divide galaxies into Seyfert and LINERs. Green points represent individual spaxels ($1''$ $\times$ $1''$), whereas the red star shows the entire $3''$ $\times$ $3''$ region. As shown in Fig. 3, all of the observed points lie in the composite region, suggesting that the central region of NGC 3665 is a mixture of star formation and a weak AGN, and is consistent with \citet{Ho1997}.

For the central $3''$ $\times$ $3''$ region, the equivalent width of H$\alpha$ ($W_{{\rm H}\alpha}$) is $\sim$3.82 {\AA}, and $\log$([{N~\sc ii}]/H$\alpha$) is about -0.21, which are also consistent with the identification of weak AGN (i.e., $\log$([{N~\sc ii}]/H$\alpha$) $>$ -0.4 and $W_{{\rm H}\alpha}$ between 3 and 6 {\AA}) in the $W_{{\rm H}\alpha}$ versus [{N~\sc ii}]/H$\alpha$ (WHAN) diagram \citep{Cid2011}. Our results are consistent with \citet{Nyland2016}, who detected two extended radio jets on scale of kilo parsecs in NGC 3665. 

Using the observed ratio of $F_{{\rm H}\alpha}/F_{{\rm H}\beta}$, we can estimate the nebular extinction, $A_{V,nebular}$ \citep{Cardelli1989}, assuming an unreddened $I_{{\rm H}\alpha}/I_{{\rm H}\beta}$ of 2.86 from \citet{Osterbrock1989}, e.g.,
 \begin{equation}
       A_{V,nebular} = 7.2 \times\ \log\
       (\frac{F_{{\rm H}\alpha}/F_{{\rm H}\beta}}{I_{{\rm H}\alpha}/I_{{\rm H}\beta}}).
   \end{equation}
We obtained that the nebular extinction, $A_{V,nebular}$, for NGC 3665 is $\sim$ 1.3, which is larger than the mean value, 1.06 $\pm$ 0.66, of 45 star-forming S0 galaxies \citep{Xiao2016} from the Sloan Digital Sky Survey \citep[SDSS;][]{Abazajian2009}.

Here we roughly calculate the surface density of star formation rate in the central $3''$ $\times$ $3''$ region, $\Sigma_{\rm SFR}$, without considering the effect of weak AGN. Using \citet{Kennicutt1998b} relation with extinction-corrected ${H_\alpha}$ luminosity: 
    \begin{equation}
        {\rm SFR}~(M_\odot~yr^{-1}) = 7.9 \times 10^{-42}~L(H_\alpha)~({\rm ergs~s^{-1}}),
    \end{equation}
    
 \noindent we derive log$\Sigma_{\rm SFR}$ $\sim$ $-0.78$ $M_{\odot}$ yr$^{-1}$ kpc$^{-2}$, which is lower than the mean value, $-$0.48 $M_{\odot}$ yr$^{-1}$ kpc$^{-2}$, of 45 star-forming S0 galaxies with the same method \citep{Xiao2016}. Since the [N~\textsc{ii}] 122 $\mu$m emission is the strongest in the center (see Fig. 2), suggesting a similar distribution of H~\textsc{ii} regions \citep{Zhao2016a}, we conclude that the star formation is concentrated in the galactic center and the rate is low.

\subsection{SED Fitting}
To better understand the infrared properties of NGC 3665, such as the total infrared luminosity ($L_{\rm TIR}$; 8-1000 $\mu$m as defined in \citealt{Sanders1996}), dust temperature ($T_{\rm dust}$) and dust mass ($M_{\rm dust}$), we used the code of Multi-wavelength Analysis of Galaxy Physical Properties\footnote{http://www.iap.fr/magphys/magphys/MAGPHYS.html} \citep[MAGPHYS;][]{da Cunha2008} to fit the observed SED of NGC 3665. Besides the {\it Herschel} PACS and SPIRE photometric results, we also compiled UV to FIR photometries from Galaxy Evolution Explorer \citep[{$GALEX$;}][]{Loubser2011}, SDSS \citep{Adelman-McCarthy2008}, Two Micron All Sky Survey \citep[2MASS;][]{Jarrett2000}, and Infrared Astronomical Satellite \citep[$IRAS$;][]{Moshir1990} catalogs. 
The measured fluxes are listed in Table 3.

MAGPHYS is a simple model to interpret in a consistent way the emission from galaxies at UV, optical and IR wavelengths in terms of their star formation histories and dust content, using a Bayesian fitting method. The library of model galaxy spectra are composed of two types of binary files: the `optical models' tracing the emission from stellar populations in galaxies, calculated using BC03 with initial mass function (IMF) from \citet{Chabrier2003} and the dust attenuation model described in \citet{Charlot2000}; the `infrared models' tracing the emission from dust, following the approach described in \citet{da Cunha2008}.
This model relies on the assumption that the total energy absorbed by dust from two main components \citep{Charlot2000}: the stellar birth clouds (star-forming regions) and the ambient diffuse ISM, and re-radiated by dust at IR wavelengths via an energy balance argument. 

We present the result of SED fitting for NGC 3665 in Fig. 4. In the top panel, the black line shows the best model fitting to the observed data (red points), the blue and red lines represent the unattenuated stellar population spectrum and the dust emission, respectively. The residuals ($L_{\rm obs} - L_{\rm mod})/L_{\rm obs}$) are shown with black squares. As we can see that the modeled spectrum is in good agreement with the observed data points from {\it GALEX} 1539 \AA\ to {\it Herschel} SPIRE 500 $\mu$m. 

Table 4 lists the derived parameters from the best-fit SED.  The dust to stellar mass ratio $M_{\rm dust}$/$M_\ast$ of NGC 3665 is about $1.1\times 10^{-4}$, which is about 3 times higher than the mean value for 39 S0+S0a galaxies observed with \emph{Herschel} \citep{Smith2012}.  As shown in \citet{Alatalo2013}, NGC 3665 has a total molecular gas mass of $\log M_{\rm gas}$ = 9.11 $\pm$ 0.01 $M_\odot$, indicating that the gas-to-dust mass ratio (GDR) is $\sim$182, which is similar to the value of the Milky Way, e.g., $\sim$120 from \citet{Li2001}, $\sim$160 from \citet{Zubko2004}, and $\sim$180 from \citet{Draine2007}. We also calculated the total infrared luminosity ($L_{\rm IR}$) of NGC 3665, integrated within 8-1000 $\mu$m from SED best-fitting model, to be $10^{9.88 \pm 0.02}$ $ L_\odot$. The error is estimated through performing a Monte Carlo simulation, sampling a series of data points according to a Gaussian distribution with the measured photometric values and errors, and repeated the same fitting procedure for 1,000 times using the simulated datasets.

We also ran another SED fitting model CIGALE\footnote{http://cigale.lam.fr} \citep{Noll2009} version 0.11.0 to make a comparison. The derived values of stellar mass and dust luminosity are $\log M_*=10.79$ $M_\odot$ and $\log L_{\rm d}=9.97$ $L_\odot$, respectively, consistent with the results from MAGPHYS. Therefore, we adopted the fitted parameters from MAGPHYS in the following analysis.

\subsection{Star Formation Rate}

To estimate the star formation rate (SFR) in NGC 3665, we adopted several different approaches: 

(1) Using the infrared and {\it GALEX} far-UV luminosity \citep{Dale2007}:
\begin{equation}
{\rm SFR}\,(M_\odot\,{\rm yr}^{-1})=4.5 \times 10^{-37}L_{\rm TIR}\,({\rm W}) + 7.1 \times 10^{-37}\nu L_\nu\,(1500{\rm\,\AA})\,({\rm W}),
\end{equation}
\noindent  where $L_{\rm TIR}$ is calculated from integration within 8-1000 $\mu$m from SED fitting (see the Section 3.2). The SFR in NGC 3665 is derived to be 1.34 $\pm$ 0.06 $M_\odot\,{\rm yr}^{-1}$. 

(2) Based only on $L_{\rm TIR}$, with the algorithm of \citet{Kennicutt1998b}: 
\begin{equation}
{\rm SFR}\,(M_\odot\,{\rm yr}^{-1})=4.5 \times 10^{-44}L_{\rm TIR}\,({\rm ergs~s^{-1}}).
\end{equation}
 The derived SFR is 1.29 $\pm$ 0.06 $M_\odot\,{\rm yr}^{-1}$. As we know, the weak AGN in NGC 3665 might contribute to the IR emission, which leads to an overestimated SFR. However, the results of CIGALE suggest that the fractional contribution of AGN to the dust emission, $f_{\rm AGN}$, is less than 0.01, thus we ignored its contribution to the infrared luminosity. 
 
(3) As shown in  \citet{Zhao2013, Zhao2016a}, the [N~\textsc{ii}] 205 $\mu$m luminosity ($L_{\rm [N\,{\scriptsize \textsc{ii}}]205\,\mu{\rm m}}$) is less affected by emissions from older stars compared to the IR luminosity, it is also a good indicator of SFR. With the 60-to-100 $\mu$m flux density ratio, $f_{60}/f_{100} \sim 0.3$ in NGC 3665, we adopted the relation suitable for the cold FIR color ($0.2\leq f_{60}/f_{100}<0.6$): 
\begin{equation}
\log\,{\rm SFR}\,(M_\odot\,{\rm yr}^{-1}) = -5.99 + \log\,L_{\rm [N\,{\scriptsize \textsc{ii}}]}\,(L_\odot). 
\end{equation}
 \noindent We calculated SFR to be 2.55$_{-1.01} ^{+1.68}$ $M_\odot\,{\rm yr}^{-1}$, where the error is estimated from the uncertainty (0.22 dex) associated with the SFR calibrator given in \citet{Zhao2016a}. 
 
(4) The [C~\textsc{ii}] 158 $\mu$m emission can also be served as a useful indicator of SFR \citep{Sargsyan2014}, as: 
\begin{equation}
\log\,{\rm SFR}\,(M_\odot\,{\rm yr}^{-1}) = \log\,L_{\rm [C\,{\scriptsize \textsc{ii}}]}\,(L_\odot) - 7.0, 
\end{equation}
 \noindent with a scatter of 0.2 dex. The derived SFR is 1.66$_{-0.97} ^{+0.61}$ $M_\odot\,{\rm yr}^{-1}$. 
 
 Therefore, SFRs from different calibrators are consistent with each other within uncertainties, and we take the averaged value of 1.7 $M_\odot\,{\rm yr}^{-1}$ as the final result.

\subsection{The Photodissociation Region}
\subsubsection{[C~\textsc{ii}] Emission from Ionized Gas}
The emission of [C~\textsc{ii}] originates from both the neutral and ionized gas, due to the low ionization potential (11.26 eV) of atomic carbon. To use the PDR models, in which only the emission from neutral gas has been taken into account, we first need to remove the [C~\textsc{ii}] emission from ionized gas. Following the method of \citet{Oberst2006,Oberst2011}, the contribution of ionized gas can be estimated with the [C~\textsc{ii}]/[N~\textsc{ii}] 205 ratio, which is only a function of electron density ($n_e$) in the H~\textsc{ii} regions after assuming a C/N abundance ratio. 
$n_e$ can be estimated with the [N~\textsc{ii}] 122/[N~\textsc{ii}] 205 ratio because of their different critical densities \citep{Oberst2006, Zhao2016a}. 
Compared the observed value ($1.95\pm 0.27$) with the theoretical curve, we determine $n_e$ = 49.5 $\pm$ 11.9 cm$^{-3}$, which is comparable to those found in ETGs and star-forming galaxies. For instance, \citet{Lapham2017} found $n_e$ = 24 cm$^{-3}$ for 11 nearby ETGs from \emph{Herschel} observations, and \citet{Diaz-Santos2017} found $n_e$ from 20 to 100 cm$^{-3}$, with the mean value of 45 cm$^{-3}$, for 240 GOALS luminous IR galaxies (LIRGs). 
In the Milky Way, the average value is measured to be 29 cm$^{-3}$ \citep{Goldsmith2015}, and in nearby spiral galaxy NGC 891, $n_e$ is ranging from 1.9 to 80 cm$^{-3}$, with a mean value of 22 cm$^{-3}$ \citep{Hughes2015}.

Based on the Lick indices (Fe5015 and Mg $b$) and single stellar population models, \citet{McDermid2015} derived the stellar metallicity at $R_{\rm e}/8$ ($\sim$ 0.87 kpc) to be [Z/H] = -0.05 $\pm$ 0.05. We calculated gas-phase metallicity based on the [{N~\sc ii}]$\lambda$6584/H$\alpha$ line ratio \citep{Kewley2002} for the same central region, to be $\sim$ 1.0 $Z_\odot$. Thus, by adopting a solar abundances of ${\rm C/H} = 1.4 \times 10^{-4}$ and N/H = 7.9 $\times$ 10$^{-5}$ from \citet{Savage1996}, we further used the derived electron density to predict the [C~\textsc{ii}]/[N~\textsc{ii}] 205 ratio in ionized gas, and then compared it with the observed values. We find that the fraction of [C~\textsc{ii}] emission from ionized gas is about 43\%. This value appears consistent with the previous results from various sources that the majority of [C~\textsc{ii}] emission comes from PDRs \citep{Abel2005,Oberst2006,Oberst2011,Farrah2013,Parkin2014,Hughes2015,Lapham2017}. After removing contribution from ionized gas for the [C~\textsc{ii}] emission, we estimated the [C~\textsc{ii}] flux originating from neutral gas is $(27.6 \pm 0.5) \times 10^{-17}$ W m$^{-2}$.

\subsubsection{Gas Heating and Cooling}
The [C~\textsc{ii}] 158 $\mu$m and [O~\textsc{i}] 63 $\mu$m are dominant coolants in neutral gas of PDRs, which can help us constrain the physical conditions of neutral ISM. The strengths of these two lines show how many interstellar UV photons heat the gas by photoelectric effect, which can be traced by emission lines during gas cooling via collisional excitation at the FIR wavelengths. Another proportion of UV photons are absorbed by dust grains, and re-emit in the infrared, which can be traced by total infrared flux. Therefore, the ratio of ([C~\textsc{ii}]+[O~\textsc{i}]63)/$F_{\rm TIR}$ is a criterion for diagnosing photoelectric heating efficiency of the interstellar gas \citep{Tielens1985}.

The total infrared flux, $F_{\rm TIR}$, is $2.19\times 10^{-13}$ W m$^{-2}$, we calculated the ([C~\textsc{ii}]+[O~\textsc{i}]63)/$F_{\rm TIR}$ ratio in PDRs to be in the range of 1.26 $\times$ 10$^{-3}$ and 1.37 $\times$ 10$^{-3}$, where the lower and upper limits were obtained by assuming zero and 3$\sigma$ fluxes of the [O~\textsc{i}] emission, respectively. Whereas the typical values of ([C~\textsc{ii}]+[O~\textsc{i}]63)/$F_{\rm TIR}$ are in the range of 10$^{-3}$ to 10$^{-2}$, both in ETGs and late-type galaxies \citep{Malhotra2001, Brauher2008, Lapham2017}. For other galaxies with spatially resolved observations, the heating efficiency varies between $\sim$$2 \times 10^{-3}$ and 10$^{-2}$ in the late spiral galaxy NGC 1097 and Seyfert 1 galaxy NGC 4559 \citep{Croxall2012}. \citet{Hughes2015} also found in NGC 891, the ([C~\textsc{ii}]+[O~\textsc{i}]63)/$F_{\rm TIR}$ ranging from $\sim$ 1 $\times$ 10$^{-3}$ to 2 $\times$ 10$^{-2}$ using \emph{Herschel} FIR spectroscopic observations. \citet{Parkin2014} showed that the heating efficiency in the disk of Centaurus A is ranging from 4 $\times$ 10$^{-3}$ to 8 $\times$ 10$^{-3}$. In the arm and inter-arm regions of M51, the average value is up to $\sim$ 10$^{-2}$, and in the nucleus decreasing to 3 $\times$ 10$^{-3}$ \citep{Parkin2013}. Therefore, NGC 3665 is among those sources having the lowest gas heating efficiency.

The low gas heating efficiency in NGC 3665 might be caused by its weak UV radiation field. The gas is mainly photoelectrically heated by the UV photons, while dust can be heated by both optical and UV photons \citep{Malhotra2000}. As shown in \citet{Abel2009}, the FIR color is a good indicator of the ionization parameter ($U$) of the ambient UV radiation field. For NGC 3665, $f_{60}/f_{100} \sim 0.3$ indicates a very small $U$ of $\sim 10^{-4}$. Therefore, there is not enough energy for the electron to collisionally excited the C$^+$ and O to higher levels.

\subsubsection{The PDR Model}
We compare IR emission line ratios to PDR model to obtain the physical properties of the PDR regions. Here we adopt the PDR model of \citet{Kaufman1999,Kaufman2006}, which has been updated based on the original model of \citet{Tielens1985}. These models assume a homogeneous semi-infinite two-dimensional slab of a PDR and solve for the chemistry, thermal balance, and radiation transfer simultaneously. For given gas-phase elemental abundances and grain properties, the model is parameterized by two free parameters: the hydrogen nucleus density, $n$, and the strength of FUV (6 eV $<$ E $<$ 13.6 eV) radiation field, $G_0$, in units of 1.6 $\times$ 10$^{-3}$ erg cm$^{-2}$ s$^{-1}$ from the local Galactic interstellar FUV field \citep{Habing1968}.

We adopt the diagnostic observed line ratios of [C~\textsc{ii}]/[O~\textsc{i}]63 versus ([C~\textsc{ii}]+[O~\textsc{i}]63)/$F_{\rm TIR}$ as mentioned in \citet{Wolfire1990}. Here we make several corrections to the observed quantities following the strategy of \citet{Zhao2016b}. 
The cloud is optically thin to the infrared continuum photon, which contributes to the actual observations from the front and back side of the cloud (especially when they are illuminated from all sides), while the models only take into account one side emission exposed to the source of UV photons. Therefore, we reduce the observed $F_{\rm TIR}$ by a factor of 2 as suggested by \citet{Kaufman1999}. 

In the PDR models, the emission line is only considered to originate from neutral gas. As mentioned above, the [C~\textsc{ii}] emission arises from both neutral and ionized gas, we first need to remove the contribution of [C~\textsc{ii}] from ionized gas, with the fraction of $\sim$43\%. Besides, we correct the geometrical effect of PDR models to the observed [C~\textsc{ii}] emission. The [C~\textsc{ii}] is marginally optically thick with optical depth $\tau$ $\sim$ 1 at the line center \citep{Kaufman1999}, thus the observed emission comes from the front side and partial back side of the cloud. We adopt the correction factor of 1.4 to divide the observed flux when comparing to the two-dimensional PDR models. More methodology details are described in \citet{Zhao2016b}. The [O~\textsc{i}]63 line is optically thick. We observe the emission only from the front side of the cloud, while the other about half of the total [O~\textsc{i}]63 emission radiates away from the line of sight. Accordingly, the actual observed [O~\textsc{i}]63 flux follows the geometrical assumption of PDR models, without any correction applied. Finally, the equation of ([C~\textsc{ii}]+[O~\textsc{i}]63)/$F_{\rm TIR}$ after correcting is listed in here: ([C~\textsc{ii}]/1.4+[O~\textsc{i}]63)/($F_{\rm TIR}$/2.0), where the [C~\textsc{ii}] emission is only taken into account originating from neutral gas.

Through a comparison of two observed line ratios to the two-dimensional PDR model, with $\chi ^{2}$ minimization in the web-based Photo Dissociation Region Toolbox \citep[PDRT;][]{Pound2008}\footnote{http://dustem.astro.umd.edu/pdrt/}, the derived values of hydrogen volume density, $n$, and the incident FUV radiation field, $G_{0}$, are listed in Table 5. We also list the best-fitting results derived from uncorrected diagnostic observed line ratios in the PDR region compared to the model. Meanwhile, both results before and after correction are shown in Fig. 5. The $G_{0}$ in NGC 3665 is significantly lower than what found in normal, star-forming, and starburst galaxies, and galaxies with strong AGN \citep{Negishi2001, Malhotra2001, Kramer2005, Oberst2011,Croxall2012,Parkin2013, Zhao2016b}. The low $G_0$ indicates a weak FUV radiation field, which is consistent with our previous analysis.

$F_{\rm TIR}$ is calculated among the whole galaxy, while [C~\textsc{ii}] emission tend to concentrate in the center of NGC 3665 (see the Fig. 2) as well as the [O~\textsc{i}]63 emission, which might result in the low $G_0$ after comparing with the PDR model. Here we focus on the central spaxel, with a size of 9$^{\prime\prime}.4$ $\times$ 9$^{\prime\prime}.4$, to discuss the physical properties of ISM. The [C~\textsc{ii}] and [O~\textsc{i}]63 line fluxes in the central region are measured using the same method as in Section 2.2.1. We calculate the central infrared luminosity of NGC 3665 using {\it Herschel} photometry at 100, 160 and 250 $\mu$m, following \citet{Galametz2013}: 
\begin{equation}
{\rm L_{TIR}}~(L_\odot\,) = (1.379 \pm 0.025)L_{100 \mu m} + (0.058 \pm 0.049)L_{160 \mu m} + (1.150 \pm 0.092)L_{250 \mu m},
\end{equation}
\noindent  where $L_{100 \mu m}$, $L_{160 \mu m}$ and $L_{250 \mu m}$ are band luminosities in the unit of $L_\odot\,$. The central infrared luminosity is estimated  to be (49.5 $\pm$ 2.3) $\times$ 10$^{7}$ $L_\odot\,$. After comparing the two observed line ratios to the PDR model, we derive two physical parameters in the central PDRs of NGC 3665: the hydrogen nucleus density, n $\sim$ 5.62 $\times$10$^{3}$ cm$^{-3}$, and the strength of FUV radiation field, G$_0$ $\sim$ 10$^{0.25}$.  The $G_{0}$ in the central region of NGC 3665 is about 3 times larger than that derived from the whole galaxy, while this value is still low enough to indicate a weak FUV radiation field.

\section{Abundant Molecular Gas and Suppressed Star Formation}

Comparing with the so-called `star-forming main sequence' (stellar mass-SFR relation) at z$\sim$0 \citep{Elbaz2007}, NGC 3665 lies 0.5 dex lower than the locus of star-forming galaxies at the fixed stellar mass, showing that star formation is suppressed. In \citet{Davis2014}, NGC 3665 has $\log \Sigma_{\rm SFR}$ down to $-2.15$ $M_{\odot}$ yr$^{-1}$ kpc$^{-2}$, and shows the largest deviation from the Kennicutt-Schmidt (KS) relation among 32 CO-detected ATLAS$^{\rm 3D}$ ETGs. The gas (atomic$+$molecular) surface density ($\Sigma_{\rm gas}$) of NGC 3665 is $\sim$ 0.9 dex larger than that of spiral galaxies at the same $\Sigma_{\rm SFR}$.  Furthermore, its molecular gas surface density \citep[2.16 $M_\odot\,{\rm pc}^{-2}$;][]{Davis2014} is significant larger than spiral galaxies at a given $\Sigma_{\rm SFR}$ in the KS plane \citep{Bigiel2008,Leroy2008,Leroy2013}. 
  \citet{Shi2011,Shi2018} proposed an extended Schmidt law, invoking the stellar mass to be a secondary role (the first is gas mass) in regulating star formation, as $\Sigma_{\rm SFR}$$\propto$($\Sigma_{\rm star}^{\rm 0.5}$$\Sigma_{\rm gas}$)$^{1.09}$.   We also compared NGC 3665 with the extended Schmidt law, and found that it has the largest offset to this relation among 20 ETGs, and has $\Sigma_{\rm star}^{\rm 0.5}$$\Sigma_{\rm gas}$ $\sim$ 1.6 dex larger than late-type galaxies at the same $\Sigma_{\rm SFR}$. Here we have calculated the stellar mass surface density to be $\log \Sigma_{\rm star}$ = 3.56 $M_\odot\,{\rm pc}^{-2}$ with the same method mentioned in the following as \citet{Fang2013}.
These results reveal that NGC 3665 has large gas reservoirs while less star formation. 

Meanwhile, NGC 3665 is in a low density environment, with the local galaxy surface density within the radius to the 10th nearest neighbor of log $\Sigma_{10} = -1.24~Mpc^{-2}$, and has been classified into field galaxy in \citet{Cappellari2011}. \citet{Young2011} further explained that the poor environment might induce in large CO storage in galaxies by cool gas accretion. We derived the contribution from cold dust to the total dust luminosity ($\xi_{\rm C}^{\rm tot}$) to be $\sim$ 70 per cent from MAGPHYS, and calculate $f_{60}/f_{100} \sim 0.3$ to classify NGC 3665 into `cold' galaxy as \citet{Zhao2016a}. These results reveal a high cool gas proportion in our source. Among current studies, scenarios proposed to prevent star formation can be roughly divided into three different routines: the cutoff of gas inflow, the removal (or heat) of cool gas in galaxies, and the stabilization of cold gas reservoirs \citep{Martig2009,Fang2013,Bluck2014}. 

We carried out a two-dimensional (2D) bulge-disk decomposition using \textsc{Galfit} \citep[version 3.0.5;][]{Peng2002,Peng2010}, and find that the SDSS \citep{Gunn1998,Aihara2011} {\em r}-band image of NGC 3665 can be fitted very well with one S\'{e}rsic component.  The fitting results are listed in Table 6, and shown in Fig. 6.  The S\'{e}rsic index is 3.81, showing that NGC 3665 is a typical bulge-dominated galaxy. \citet{Fang2013} found the stellar mass surface density within 1 kpc, $\Sigma_1$, is a critical indicator of the star formation suppression. We calculate $\Sigma_1$ following Fang's method. Using the relation between stellar mass-to-{\em i}-band luminosity ratio  and rest-frame {\em g}$-${\em i} color, log $M/L_{i} = 1.15 + 0.79~(g-i)$, we measure the value of log $\Sigma_1$ to be $\sim$ 9.79 $M_\odot\,{\rm kpc}^{-2}$. This value is about 0.15 dex higher than the best-fit $\Sigma_1$ vs. stellar mass relation for green valley and red sequence galaxies at the fixed stellar mass \citep[Figure 4 in][]{Fang2013}, indicating a relatively compact spheroidal stellar component in our source. With such a compact, massive bulge to stabilize cold gas reservoirs, star formation can be suppressed effectively in NGC 3665.

On the other hand, we estimate the Toomre $Q$ parameter to explore whether the molecular gas disk is stable enough to against the gravitational fragmentation, following the criteria of \citet{Toomre1964}:
 \begin{equation}
     Q_{\rm gas} =  \frac{\sigma_{\rm gas}{\kappa}}{{\pi}G{\Sigma_{\rm gas}}} > 1.
   \end{equation}
\noindent  $\sigma_{\rm gas}$ is the molecular gas velocity dispersion to be 12.53 km s$^{-1}$ with the CARMA observations \citep{Onishi2017}, $\kappa$ is the epicyclic frequency adopted of a approximate relation $\kappa$ $\sim$$\sqrt{2}$ $V$(R)/R, G is Newton's gravitational constant, and $\Sigma_{\rm gas}$ is the surface density of the gas disk to be $\sim$145 $M_\odot\,{\rm pc}^{-2}$. Thus we derive $Q_{\rm gas}$ $\sim$ 2.6, suggesting a stable molecular gas disk to against large-scale gas self-gravitational collapse.

Besides the effect of a massive bulge, this galaxy has two extended jets on kilo parsec scales \citep{Nyland2016}, which can both heat the halo gas and expel a fraction of the cool gas \citep[AGN `radio-mode' feedback;][]{Croton2006,Ogle2007,Nesvadba2010,Lanz2016,Smethurst2016,Combes2017}. In addition, on account of the high stellar mass of 10.79 $M_\odot\,$, there is another possible mechanism to prevent star formation by shock heating the gas inflow from the halo \citep{keres2005,Dekel2006,Cappellari2016}. However, with the low percentage of hot gas, these two modes of star formation suppression are unlikely the dominant mechanisms in NGC 3665. The metallicity is $\sim$ 1.0 $Z_\odot$, thus we can exclude the effect of the metallicity in lowing star formation \citep{Shi2014}.

Consequently, the suppression of star formation in NGC 3665 is most possibly caused by its compact, massive bulge through stabilizing cold gas, which enables NGC 3665 to serve as a good observational sample for the stabilization of cold gas reservoirs, and is influenced by the `radio-mode' feedback as well as the virial shocks. The low rate of star formation and weak AGN produce a weak UV radiation field (shown details in section 3.4.3), thus more dust is heated by old stars with inefficient photoelectric heating of the gas. The weak UV radiation field can not produce much [C~\textsc{ii}] and [O~\textsc{i}]63 emission and lead to somewhat low ([C~\textsc{ii}]+[O~\textsc{i}]63)/$F_{\rm TIR}$ in this atypical early-type galaxy.

\section[]{Conclusions}
We present \emph{Herschel} FIR photometric and spectroscopic observations of NGC 3665. To better understand the nuclear activity in NGC 3665, we also conducted optical spectroscopic observations. By combining the multi-wavelength data from literature and fitting the observed SED, we obtain dust luminosity, stellar and dust mass, dust temperature, infrared luminosity, and gas-to-dust mass ratio in NGC 3665. We discuss gas heating and cooling efficiency in the PDR regions and compare observed emission line ratios to the \citet{Kaufman1999,Kaufman2006} PDR models to derive hydrogen nucleus density and strength of FUV radiation field.  The main results are summarized as follows: 

\begin{enumerate}
 
 \item From the PACS spectroscopic maps of [C~\textsc{ii}] 158 $\mu$m and [N~\textsc{ii}] 122 $\mu$m, we find that both neutral and ionized gas have extend structures and follow the CO($1-0$) gas disk distribution. The fluxes are strongest at the center, and gradually weaker outwards. 

 \item NGC 3665 has dust-to-stellar mass ratio $M_{\rm dust}/M_\ast \sim 1.1 \times 10^{-4}$, which is nearly 3 times larger than the mean value of local S0+S0a galaxies. The gas-to-dust mass ratio is 182, similar to that in the Milky Way, indicating a large gas reservoir.
 
 \item According to the BPT diagnostic diagram, NGC 3665 contains both star formation and a weak AGN in the central region. We calculated the SFR to be around 1.7 $M_\odot\,{\rm yr}^{-1}$ based on several different methods. 

 \item The electron density of ionized gas in NGC 3665, based on the [N~\textsc{ii}] 122/[N~\textsc{ii}] 205 ratio, is $n_e = 49.5 \pm 11.9$ cm$^{-3}$. The contribution of ionized gas region to the total [C~\textsc{ii}] emission is about 43\%, which is consistent with the previous results that the majority of [C~\textsc{ii}] emission comes from PDRs.

 \item The ([C~\textsc{ii}]+[O~\textsc{i}]63)/$F_{\rm TIR}$ line ratio is in the range of 1.26 $\times$ 10$^{-3}$ and 1.37 $\times$ 10$^{-3}$, indicating that NGC 3665 almost has the lowest gas heating efficiency in PDRs among different kinds of galaxies.
 
 \item A comparison between the observed emission line ratios and the theoretical PDR models gives that the hydrogen nucleus density $n \sim 10^{3.75}$ cm$^{-3}$, and the strength of FUV radiation field $G_0\sim 10^{-0.25}$, indicating a very weak UV radiation field in NGC 3665.

 \item After comparing our results with previous works, we find that NGC 3665 has large gas reservoirs while low-level star formation. The suppressed star formation is most possibly caused by its compact, massive bulge through stabilizing cool gas reservoirs.

\end{enumerate}

\section*{Acknowledgments}
The authors are very grateful to the anonymous referee for critical comments and instructive suggestions, which significantly strengthened the analyses in this work. We thank Dr. Daizhong Liu and Dr. Zhiyu Zhang for helpful guidances about data reductions for \emph{Herschel} photometric and spectroscopic observations, thank Peng Wei for running SED fitting model CIGALE to help us compare the results with those in model MAGPHYS, and thank David Elbaz, Yifei Jin and Longji Bing for valuable discussions and advices which improved this paper.
We are grateful to Sebasti\'an S\'{a}nchez, resident astronomer at CAHA, for the optical spectroscopic observations and Rub\'en Garc\'ia Benito for his help of data reduction. 
This work is supported by the National Key Research and Development Program of China (No. 2017YFA0402703 and 2017YFA0402704), and by the National Natural Science Foundation of China (Nos. 11673057 and 11733002).

PACS has been developed by a consortium of institutes led by MPE (Germany) and including UVIE (Austria); KU Leuven, CSL, IMEC (Belgium); CEA, LAM (France); MPIA (Germany); INAF-IFSI/OAA/OAP/OAT, LENS, SISSA (Italy); IAC (Spain). This development has been supported by the funding agencies BMVIT (Austria), ESA-PRODEX (Belgium), CEA/CNES (France), DLR (Germany), ASI/INAF (Italy), and CICYT/MCYT (Spain).  SPIRE has been developed by a consortium of institutes led by Cardiff University (UK) and including Univ. Lethbridge (Canada); NAOC (China); CEA, LAM (France); IFSI, Univ. Padua (Italy); IAC (Spain); Stockholm Observatory (Sweden); Imperial College London, RAL, UCL-MSSL, UKATC, Univ. Sussex (UK); and Caltech, JPL, NHSC, Univ. Colorado (USA). This development has been supported by national funding agencies: CSA (Canada); NAOC (China); CEA, CNES, CNRS (France); ASI (Italy); MCINN (Spain); SNSB (Sweden); STFC and UKSA (UK); and NASA (USA).  HIPE is a joint development by the Herschel Science Ground Segment Consortium, consisting of ESA, the NASA Herschel Science Center, and the HIFI, PACS and SPIRE consortia. 
Based on observations collected at the Centro Astron\'omico Hispano Alem\'an (CAHA) at Calar Alto, operated jointly by the Max-Planck Institut f\"ur Astronomie and the Instituto de Astrof\'\i sica de Andaluc\'\i a (CSIC).

\newpage
\begin{deluxetable}{lccccc}\label{properties}
\tablecolumns{6} \tablewidth{0pc} \tabletypesize{\footnotesize}
\tablecaption{A summary of the \emph{Herschel} PACS and SPIRE Photometric Observations}
\tablehead{ \colhead{Band} & \colhead{Beam FWHM Size} & \colhead{Pixel Size} & \colhead{Color Correction} & \colhead{Aperture Correction} & \colhead{Flux} \\
\colhead{(${\mu m}$)} & \colhead{(arcsec)} & \colhead{(arcsec)} & \colhead{} & \colhead{} & \colhead{(Jy)}} \startdata

100 & 7.7 & 1.6 & 1.000 & 1.271 & $6.68 \pm 0.34$ \\ 
160 & 12 & 3.2 & 1.004 & 1.271 & $7.27 \pm 0.37$ \\ 
250 & 18.1 & 6 & 0.9121 & 1.256 & $2.84 \pm 0.45$ \\ 
350 & 24.9 & 10 & 0.9161 & 1.256 & $1.09 \pm 0.12$ \\ 
500 & 36.4 & 14 & 0.9005 & 1.256 & $0.35 \pm 0.10$ \\
\enddata
\\
\end{deluxetable}

\begin{deluxetable}{lccccccc}\label{properties}
\tablecolumns{8} 
\tablewidth{0pt} 
\tabletypesize{\scriptsize}
\tablecaption{Fine-structure Lines Observed with the PACS and SPIRE Spectrometer}
\tablehead{ \colhead{Line} & \colhead{$\lambda$} & \colhead{ObsID} & \colhead{Obs Date} & \colhead{Spec. Resolution} & \colhead{Angular Resolution}& \colhead{Flux}& \colhead{FWHM} \\
 & \colhead{($\mu m$)} &  & \colhead{} & \colhead{(km s$^{-1}$)} & \colhead{(arcsec)}& \colhead{(10$^{-17}$ W m$^{-2}$)} & \colhead{(km s$^{-1}$)}} \startdata

\mbox{[{O~\sc i}]$^{3}$P$_{1}$--$^{3}$P$_{2}$} & 63.18 & 1342223367 & 2011 Jun 30 & $\sim$86  & $\sim$9.4 & $<$ 2.48 & \nodata \\ 
\mbox{[{N~\sc ii}]$^{3}$P$_{2}$--$^{3}$P$_{1}$} & 121.90 & 1342234055 & 2011 Dec 11 & $\sim$290  & $\sim$10 & $14.09 \pm 1.98$ & 279, 221 \\ 
\mbox{[{C~\sc ii}]$^2$P$_{3/2}$--$^2$P$_{1/2}$} & 157.74 & 1342234055 & 2011 Dec 11 & $\sim$238  & $\sim$11.5 & $48.09 \pm 0.84$ & 248, 251\\ 
\mbox{[{N~\sc ii}]$^{3}$P$_{1}$--$^{3}$P$_{0}$} & 205.18 & 1342247121 & 2012 Jun 18 & $\sim$297 & $\sim$17 & $7.24 \pm 0.10$ & 568\\ 
\mbox{[{C~\sc i}]$^{3}$P$_{2}$--$^{3}$P$_{1}$} & 370.42 & 1342247121 & 2012 Jun 18 & $\sim$536 & $\sim$36 & $ 0.40 \pm 0.05$ & \nodata\\
\enddata
\\
\tablecomments{Column (1): atomic fine-structure line; columns (2): wavelength; columns (3): observation ID; column (4): observation date; column (5) and (6): spectral resolution and angular resolution from the PACS Observer's Manual and SPIRE Handbook; column (7): measured fluxes of each atomic fine-structure line for NGC 3665; column (8): intrinsic line width.}

\end{deluxetable}

\newpage
\begin{deluxetable}{lccccccc}\label{properties}
\tablecolumns{8} \tablewidth{0pc} \tabletypesize{\scriptsize}
\tablecaption{Photometry}
\tablehead{ \colhead{Properties} & \colhead{FUV} & \colhead{NUV} & \colhead{\em u} & \colhead{\em g} & \colhead{\em r} & \colhead{\em i} & \colhead{\em z} \\
\colhead{(1)} & \colhead{(2)} & \colhead{(3)} & \colhead{(4)} & \colhead{(5)} & \colhead{(6)}& \colhead{(7)}& \colhead{(8)}} \startdata

Flux (mJy) & $0.28 \pm 0.02$ & $1.07 \pm 0.02$ & $11.10 \pm 0.03$ & $58.50 \pm 0.11$ & $123.00 \pm 0.23$ & $187.00 \pm 0.35$ & $263.00 \pm 0.48$\\ 
\hline
\\

\hline
\hline
\\
 \em J & \em H & \em K$_s$ & $I12_{\mu m}$ & $I25_{\mu m}$ & $I60_{\mu m}$& $I100_{\mu m}$ &\\
(9) & (10) & (11) & (12) & (13) & (14)& (15) &\\
\hline

 $567.00 \pm 4.72$ & $692.00 \pm 7.04$ & $564.00 \pm 7.32$ & $112.00 \pm 20.10$ & $162.00 \pm 21.20$ & $1920.00 \pm 115.00$ & $6340.00 \pm 317.00$ &

\enddata
\\
\tablecomments{Column (1): flux in unit of mJy; columns (2)-(3): flux of the {\it GALEX} far-UV (FUV) and near-UV (NUV) band, respectively; columns (4)-(8): flux of the SDSS {\em u}, {\em g}, {\em r}, {\em i}, and {\em z} band, respectively; columns (9)-(11): flux of the 2MASS {\em J}, {\em H}, and {\em K$_s$} band, respectively; columns (12)-(15): fluxes of the four {\it IRAS} bands.}
\end{deluxetable}

\begin{deluxetable}{lccccccc}\label{properties}
\tablecolumns{8} 
\tablewidth{0pt} 
\tabletypesize{\footnotesize}

\tablecaption{Derived properties from SED Fitting}
\tablehead{ \colhead{$\log M_\ast$} & \colhead{$\log L_{\rm d}$} & \colhead{$\log M_{\rm d}$} & \colhead{$T_{\rm W}^{\rm BC}$} & \colhead{$T_{\rm C}^{\rm ISM}$} & \colhead{$\log M_{\rm gas}$}& \colhead{$M_{\rm gas}/M_{\rm d}$}& \colhead{$\log L_{\rm TIR}$} \\

 \colhead{($M_\odot$)} & \colhead{($L_\odot$)} & \colhead{($M_\odot$)}  & \colhead{(K)} & \colhead{(K)} & \colhead{($M_\odot$)}&  & \colhead{($L_\odot$)}} \startdata

10.79$^{+0.07}_{-0.02}$ & 9.90$^{+0.01}_{-0.03}$ & 6.85$^{+0.04}_{-0.05}$ & 55.81$^{+3.65}_{-9.05}$ & 22.86$^{+0.50}_{-0.33}$ & 9.11 $\pm$ 0.01\tablenotemark{a} & 182 & 9.88 $\pm$ 0.02\tablenotemark{b}\\ 
\enddata
\tablecomments{The column from left to right are: stellar mass, dust luminosity, dust mass, temperature of warm dust component in stellar birth clouds, temperature of cold dust component in diffuse ISM, total molecular gas mass, ratio of gas mass to dust mass, and total infrared luminosity. Uncertainties from MAGPHYS are the 16th-84th percentile range of the likelihood distribution from SED fitting.}
\tablenotetext{a}{The total molecular gas mass is from the CARMA observations \citep{Alatalo2013}.}
\tablenotetext{b}{The total infrared luminosity is integrated from SED best-fitting model between 8 and 1000 $\mu$m, with the error obtained from a Monte Carlo simulation.}
\end{deluxetable}

\newpage
\begin{deluxetable}{lcc}\label{properties}
\tablecolumns{3} \tablewidth{0pc} \tabletypesize{\footnotesize}
\tablecaption{Results from the PDR model}
\tablehead{ \colhead{Case} & \colhead{$\log n$} & \colhead{$\log G_{0}$}  \\
\colhead{} & \colhead{(cm$^{-3}$)} & \colhead{(1.6 $\times$ 10$^{-3}$ erg cm$^{-2}$ s$^{-1}$)} } \startdata

Uncorrected\tablenotemark{a} & 3.75 & -0.25 \\ 
Corrected\tablenotemark{b} & 4.00 & -0.25 

\enddata

\tablenotetext{a}{The uncorrected values are derived from the best fit, including the observed [C~\textsc{ii}] only from the neutral gas region, all observed [O~\textsc{i}]63 emission and $F_{\rm TIR}$.}
\tablenotetext{b}{The corrected values contain the [C~\textsc{ii}] divided by a factor of 1.4 in the neutral gas region and the $F_{\rm TIR}$ reduced by 2.0.}
\end{deluxetable}

\newpage
\begin{deluxetable}{lcccccc}\label{properties}
\tablecolumns{7} \tablewidth{0pc} \tabletypesize{\footnotesize}
\tablecaption{Results from the {\em r}-band bulge-disk decomposition}
\tablehead{ \colhead{Component} & \colhead{$m_r$} & \colhead{$R_{e}$} & \colhead{S\'{e}rsic index} & \colhead{$b/a$} & \colhead{PA} & \colhead{chi$^{2}/$nu}  \\
\colhead{} & \colhead{(mag)} & \colhead{(arcsec)}& \colhead{} & \colhead{}& \colhead{(deg)}} \startdata

one S\'{e}rsic & 11.18 & 50.15 & 3.81 & 0.78 & 28.12 & 0.95 \\

\enddata
\end{deluxetable}

\newpage
\begin{figure}[hbt]
  \centering
  \includegraphics[width=17cm]{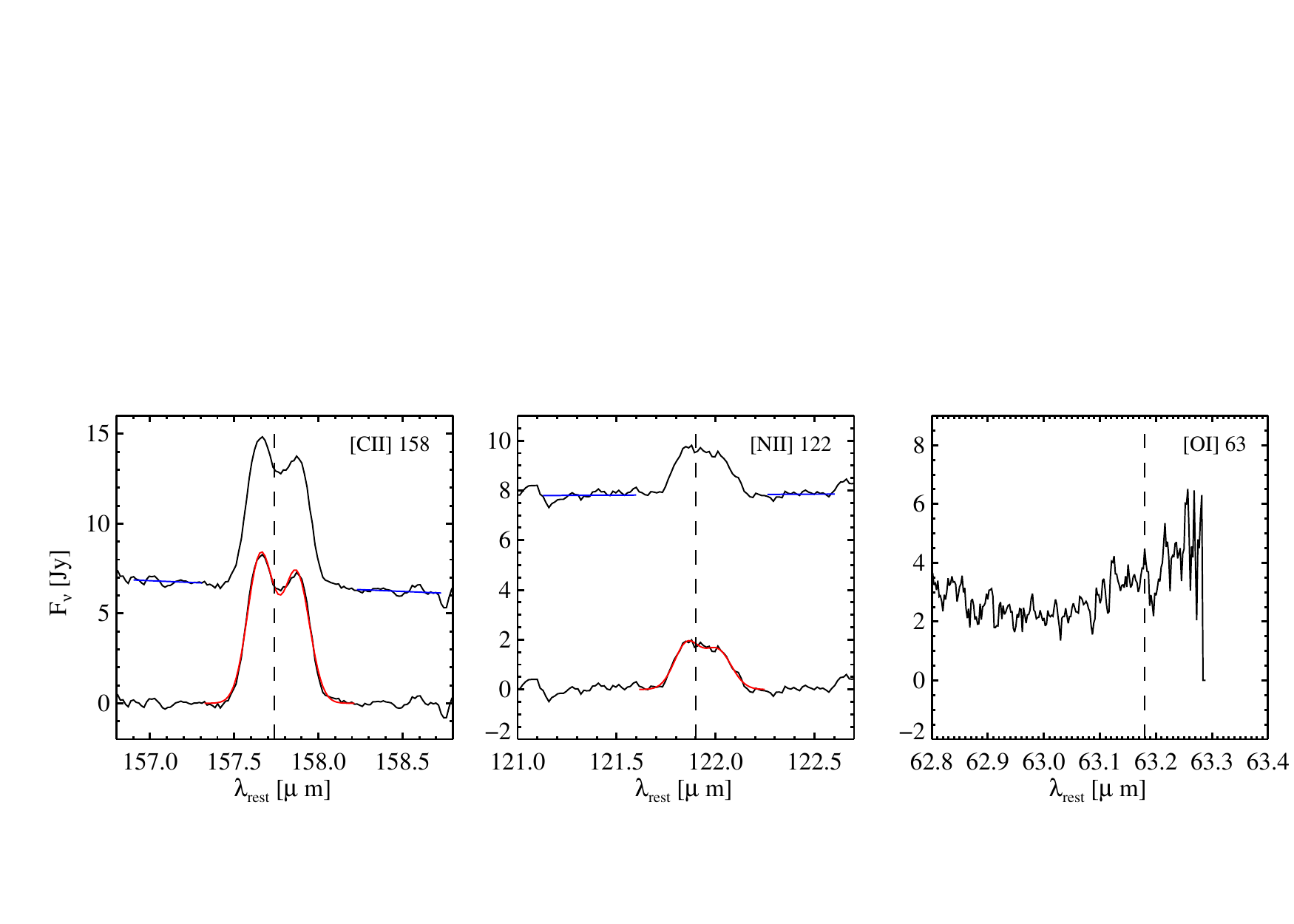}\\
  \caption{The [C~\textsc{ii}] 158 $\mu$m, [N~\textsc{ii}] 122 $\mu$m, and [O~\textsc{i}] 63 $\mu$m spectra combined within central 3$\times$3 spaxels. In the left two panels, the top and bottom black lines show observed spectra and emission lines with two Gaussian components fitted in red, respectively. The continuum emission was fitted by first-order polynomial in blue. The right panel shows observed spectra of [O~\textsc{i}] 63 $\mu$m.}\label{three_spec_fig}
\end{figure}

\newpage
\begin{figure}[hbt]
  \centering
  \includegraphics[width=17cm]{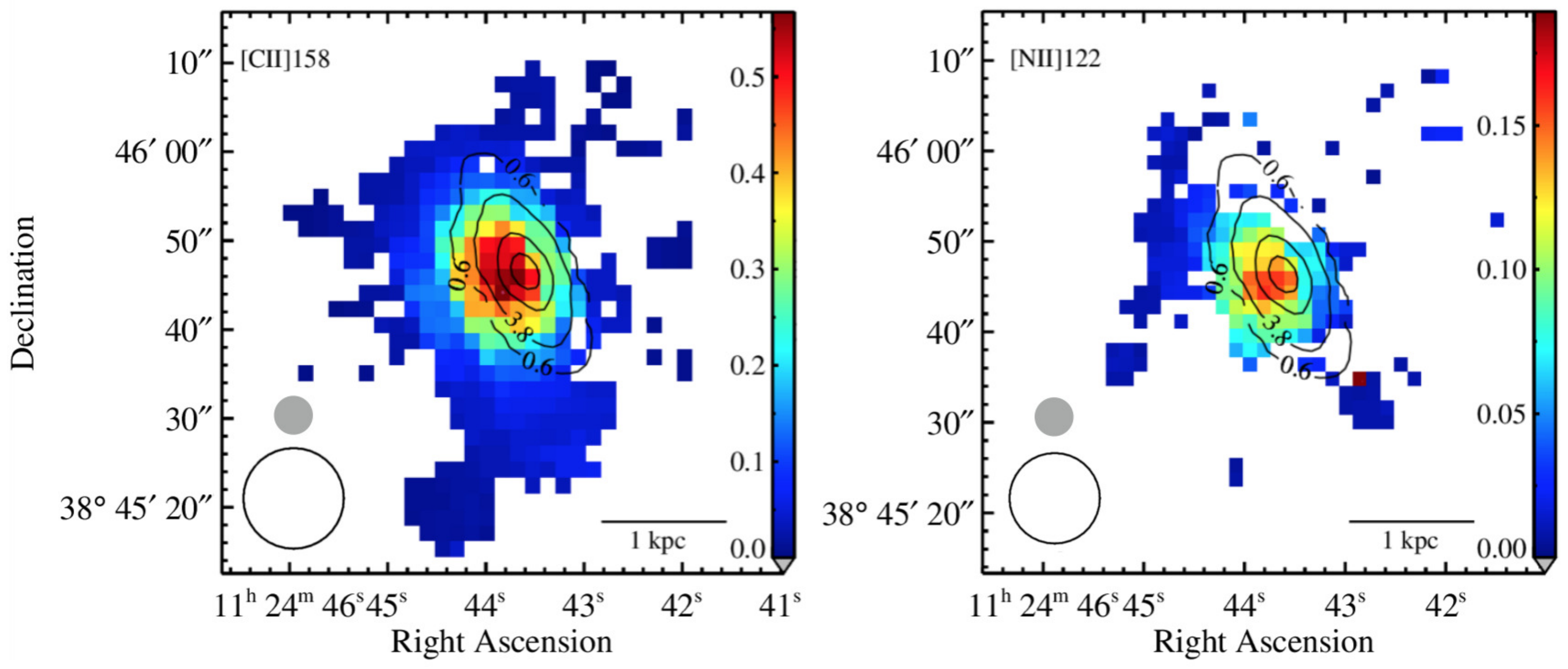}\\
  \caption{The [C~\textsc{ii}] 158 $\mu m$ and [N~\textsc{ii}] 122 $\mu$m integrated intensity maps (color scale), overlaid with contours of CO($1-0$) integrated intensity (moment0) from CARMA. We have used the 3$\sigma$ cut to highlight the robust detections. Contour levels are 2.5, 16, 50, 84 per cent of the peak, while the peak flux is 24.04 Jy beam$^{-1}$ km s$^{-1}$. The color table on the right of each panel provides the integrated flux scale of [C~\textsc{ii}] 158 $\mu m$ and [N~\textsc{ii}] 122 $\mu$m, respectively, with the unit in 10$^{-17}$ W m$^{-2}$. The synthesized beam of \emph{Herschel} and CARMA are shown in the bottom-left corner, with black open circle and grey filled circle, respectively. North is up, and east is to the left.}\label{contour_two_fig}
\end{figure}

\newpage
\begin{figure}[hbt]
  \centering
  \includegraphics[width=17cm]{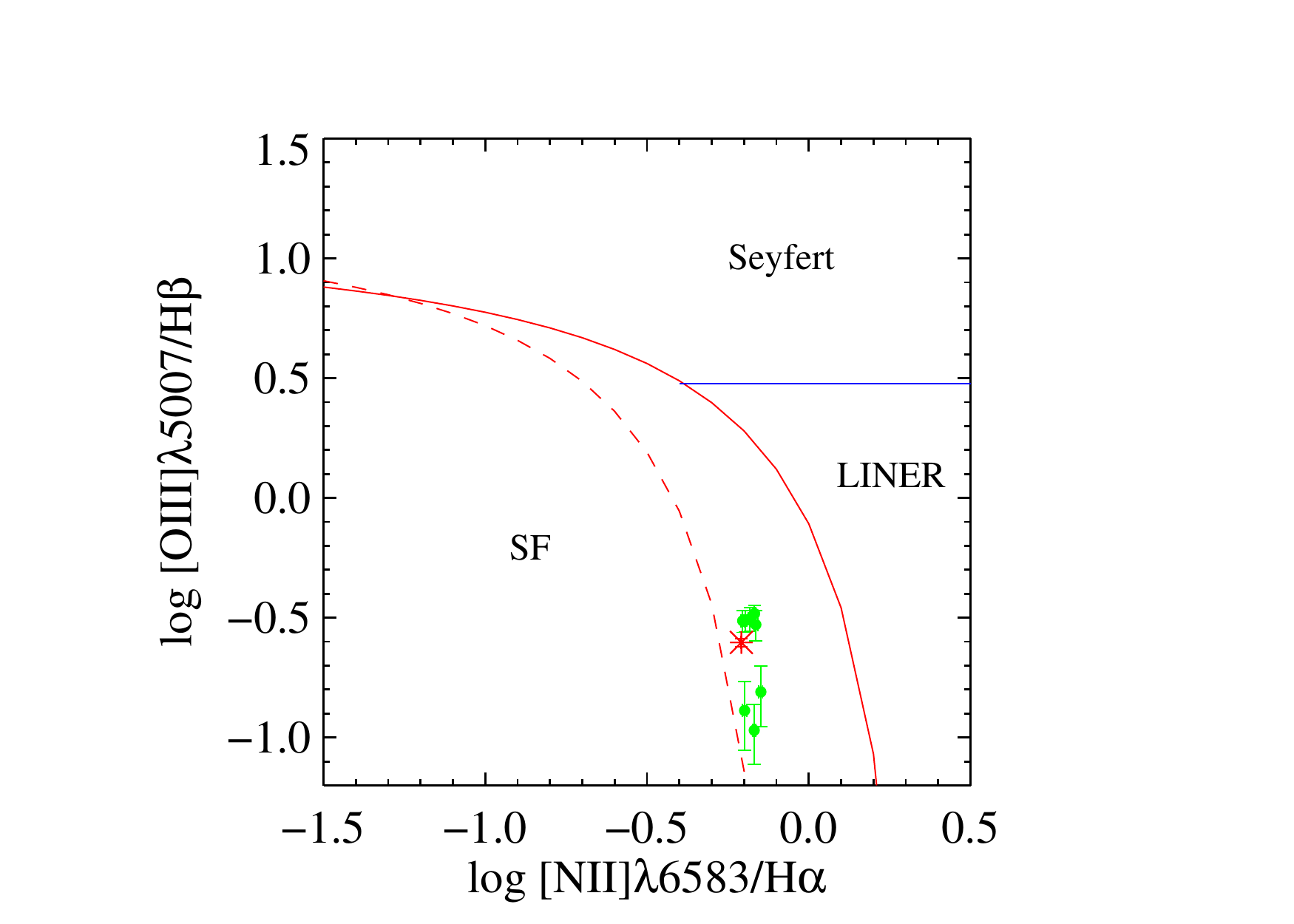}\\
  \caption{The BPT diagram in the central $3''$ $\times$ $3''$ spaxels. Green points are each individual spaxel, which lie between the red solid line \citep{Kewley2001} and the red dash line and blue line \citep{Kauffmann2003}. Red star shows 9 spaxels as a whole. Both the green points and red star, with the uncertainties, lie in the composite region in this diagnostic diagram.}\label{BC03_fig}
\end{figure}

\newpage
\begin{figure}[hbt]
  \centering
  \includegraphics[width=17cm]{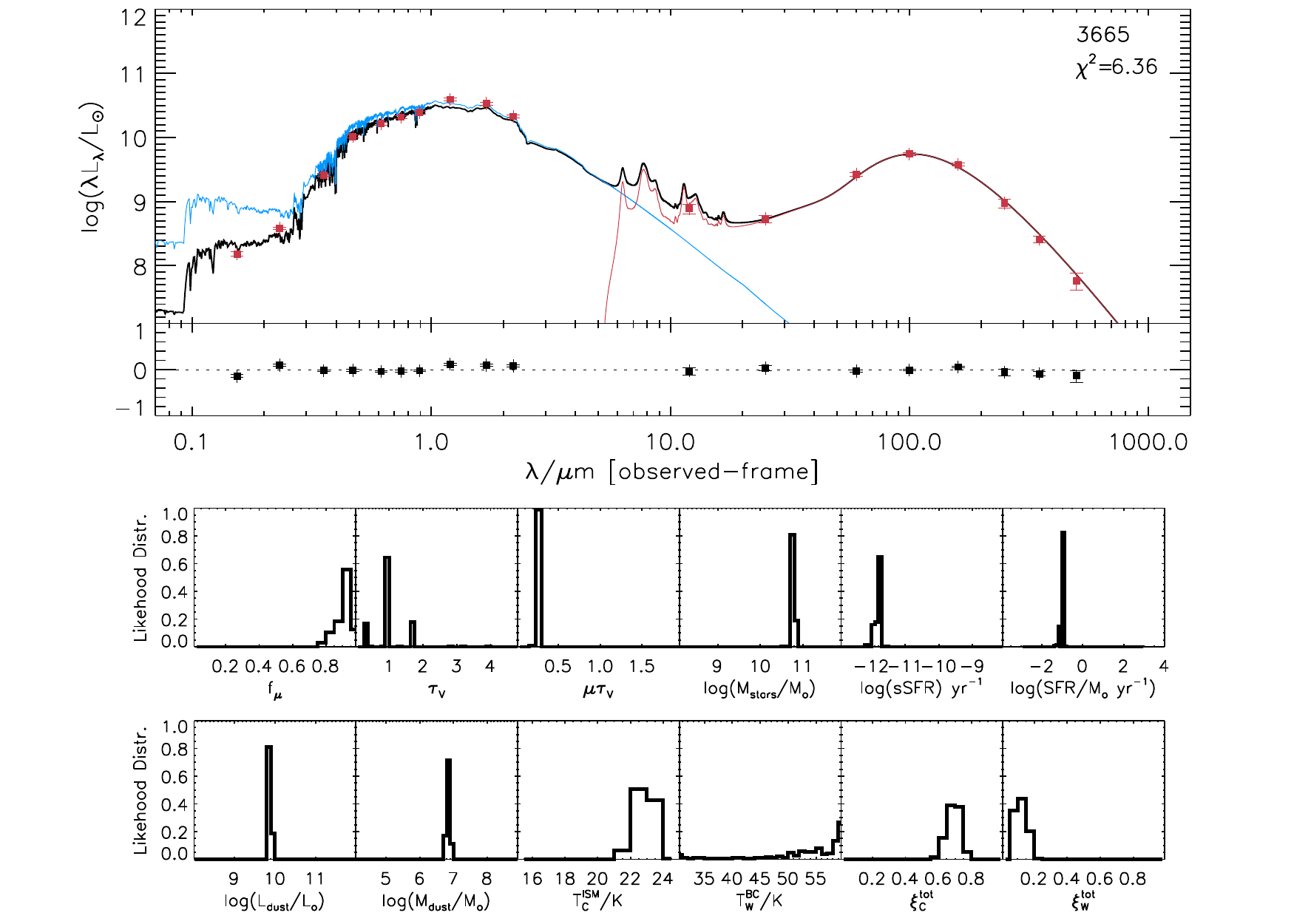}\\
  \caption{Best MAGPHYS model fits (black line) to the observed SED (red points) of the NGC 3665. The data are composed of two {\it GALEX}, five SDSS, three 2MASS, three {\it IRAS}, two PACS, and three SPIRE bands. In the top panel, the blue line represents the unattenuated stellar population spectrum, and the red line represents the dust emission. For each observational point, the red error bar indicates the measurement error, which also shown with the residuals ($\rm L_{obs} - \rm L_{mod})$/$\rm L_{obs}$ in black. The bottom twelve minor panels show the likelihood of physical parameters derived from fits to the observed spectral energy distribution, including the fraction of the total dust luminosity accounted by dust in the diffuse ISM ($f_\mu$), the total {\em V}-band optical depth of the dust ($\tau_V$), the {\em V}-band optical depth of the dust in the diffuse ISM ($\mu$$\tau_V$), stellar mass, specific star formation rate, star formation rate, total stellar luminosity absorbed by dust, the equilibrium temperature of cold dust in the diffuse ISM ($T_{\rm C}^{\rm ISM}$), the equilibrium temperature of warm dust in the stellar birth clouds ($T_{\rm W}^{\rm BC}$), the fractional contribution by cold dust to the total dust luminosity ($\xi_{\rm C}^{\rm tot}$), and the fractional contribution by warm dust to the total dust luminosity ($\xi_{\rm W}^{\rm tot}$).}\label{sed_fig}
\end{figure}

\newpage
\begin{figure}[hbt]
  \centering
  \includegraphics[width=8cm]{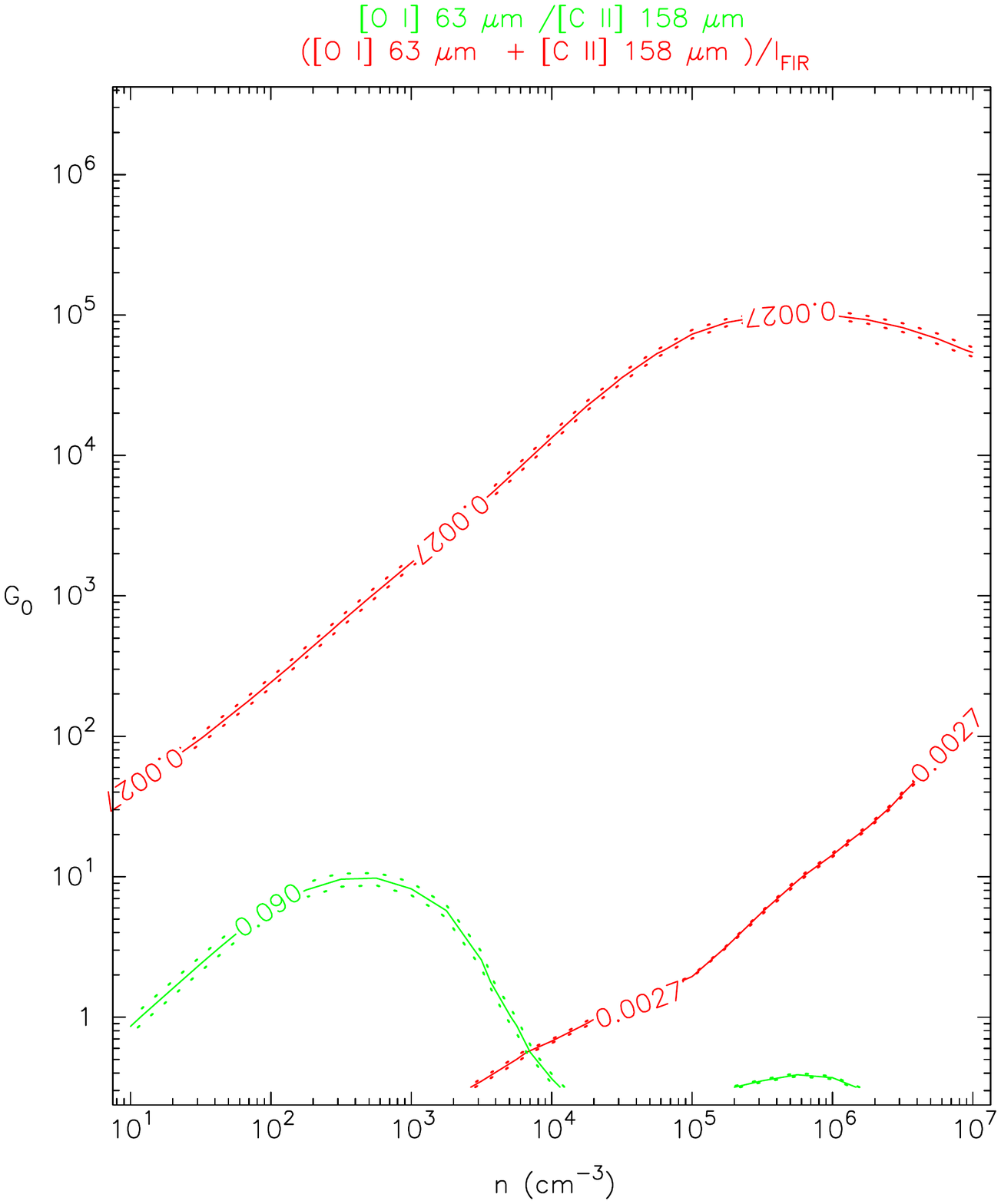}
    \includegraphics[width=8cm]{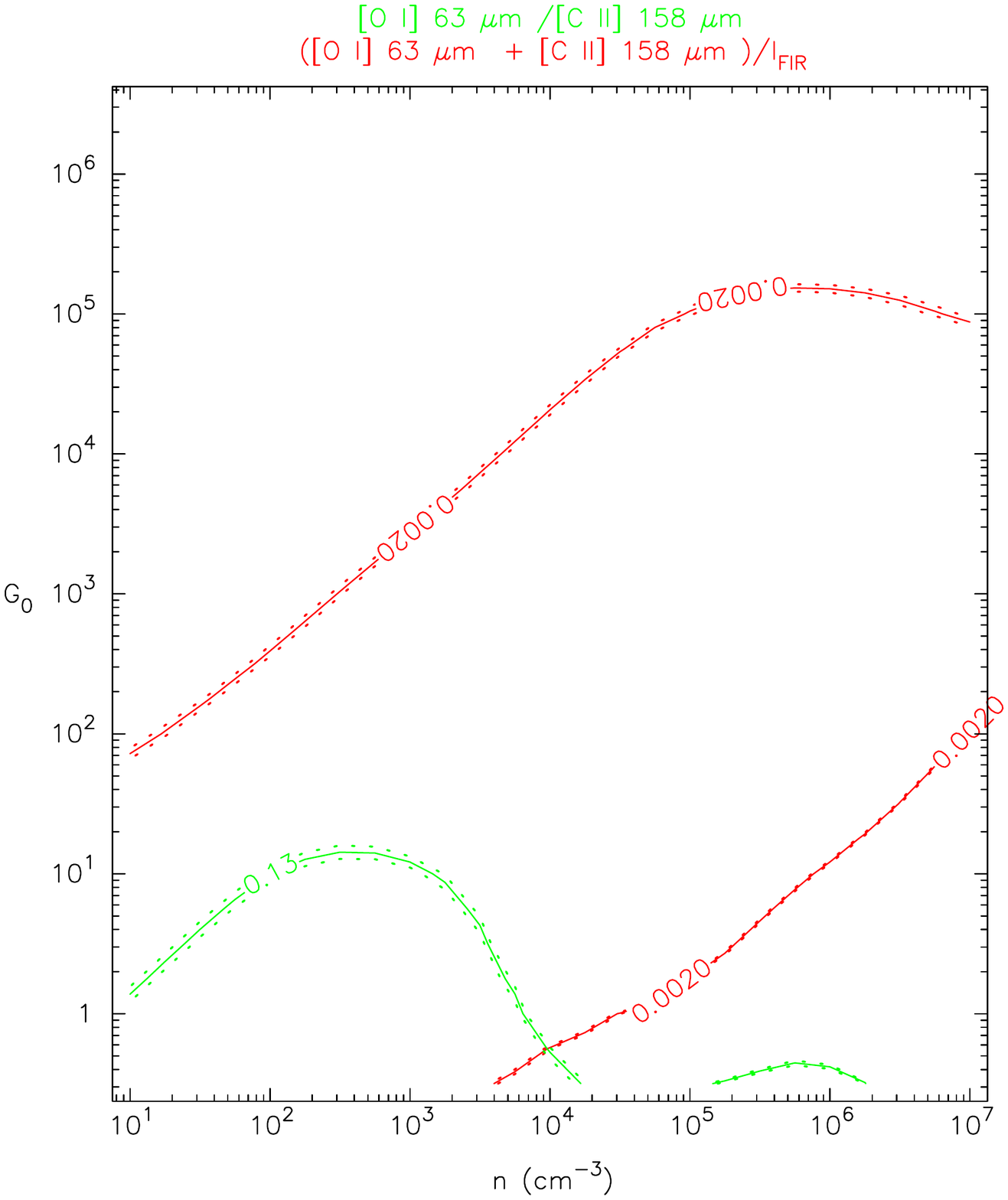}
  \caption{The line ratio curves created from on-line PDRT, as a function of the hydrogen nucleus density, n, and the strength of FUV radiation field, G$_0$. Solid lines represent computed ratios, and the dashed lines represent error bar. The intersections of two ratio curves indicate the best-fit quantities. The left/right panels are corresponding to the uncorrected/corrected results, respectively, in Table 5.}\label{PDR_fig}
\end{figure}

\newpage
\begin{figure}[hbt]
  \centering
  \includegraphics[width=17cm]{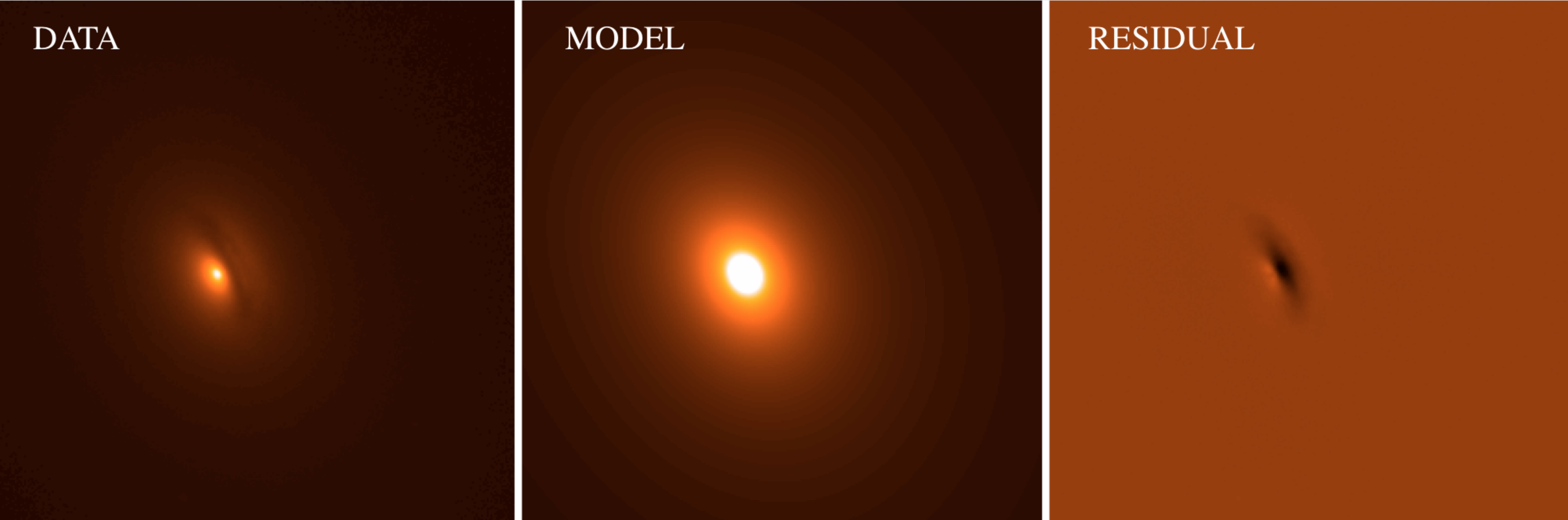}\\
  \caption{The 2D bulge-disk decomposition of NGC 3665 with \textsc{Galfit}. Panels from left to right are r-band observed image, model image, and model-subtracted residual image. The black region in residual image shows a dust structure near the galaxy center, which has been masked out before the fitting.}\label{sed_fig}
\end{figure}


\begin{thebibliography}{}

\bibitem[Abazajian et al.(2009)]{Abazajian2009} Abazajian, K.~N., 
Adelman-McCarthy, J.~K., Ag{\"u}eros, M.~A., et al.\ 2009, \apjs, 182, 543 
\bibitem[Abel et al.(2005)]{Abel2005} Abel, N.~P., Ferland, G.~J., Shaw, G., \& van Hoof, P.~A.~M.\ 2005, \apjs, 161, 65 
\bibitem[Abel et al.(2009)]{Abel2009} Abel, N.~P., Dudley, C., Fischer, J., Satyapal, S., \& van Hoof, P.~A.~M.\ 2009, \apj, 701, 1147 
\bibitem[Adelman-McCarthy et al.(2008)]{Adelman-McCarthy2008} Adelman-McCarthy, J.~K., Ag{\"u}eros, M.~A., Allam, S.~S., et al.\ 2008, \apjs, 175, 297-313 
\bibitem[Aihara et al.(2011)]{Aihara2011} Aihara, H., Allende Prieto, C., An, D., et al.\ 2011, \apjs, 193, 29 
\bibitem[Alatalo et al.(2013)]{Alatalo2013} Alatalo, K., Davis, T.~A., Bureau, M., et al.\ 2013, \mnras, 432, 1796 
\bibitem[Baldwin, Phillips \& Terlevich (1981)]{Baldwin1981} Baldwin, J.~A., Phillips, M.~M., \& Terlevich, R.\ 1981, \pasp, 93, 5 
\bibitem[Balog et al.(2014)]{Balog2014} Balog, Z., M{\"u}ller, T., Nielbock, M., et al.\ 2014, Experimental Astronomy, 37, 129 
\bibitem[Bigiel et al.(2008)]{Bigiel2008} Bigiel, F., Leroy, A., Walter, F., et al.\ 2008, \aj, 136, 2846 
\bibitem[Birnboim \& Dekel(2003)]{Birnboim2003} Birnboim, Y., \& Dekel, A.\ 2003, \mnras, 345, 349 
\bibitem[Bluck et al.(2014)]{Bluck2014} Bluck, A.~F.~L., Mendel, J.~T., Ellison, S.~L., et al.\ 2014, \mnras, 441, 599 
\bibitem[Brauher et al.(2008)]{Brauher2008} Brauher, J.~R., Dale, D.~A., \& Helou, G.\ 2008, \apjs, 178, 280-301 
\bibitem[Brinchmann et al.(2004)]{Brinchmann2004} Brinchmann, J., Charlot, S., White, S.~D.~M., et al.\ 2004, \mnras, 351, 1151 
\bibitem[Bruzual \& Charlot(2003)]{BC2003} Bruzual, G., \& Charlot, S.\ 2003, \mnras, 344, 1000 
\bibitem[Cappellari et al.(2011)]{Cappellari2011} Cappellari, M., Emsellem, E., Krajnovi{\'c}, D., et al.\ 2011, \mnras, 413, 813 
\bibitem[Cappellari(2016)]{Cappellari2016} Cappellari, M.\ 2016, \araa, 54, 597 
\bibitem[Cardelli et al.(1989)]{Cardelli1989} Cardelli, J.~A., Clayton, G.~C., \& Mathis, J.~S.\ 1989, \apj, 345, 245 
\bibitem[Chabrier(2003)]{Chabrier2003} Chabrier, G.\ 2003, \pasp, 115, 763
\bibitem[Charlot \& Fall(2000)]{Charlot2000} Charlot, S., \& Fall, S.~M.\ 2000, \apj, 539, 718 
\bibitem[Cid Fernandes et al.(2011)]{Cid2011} Cid Fernandes, R., Stasi{\'n}ska, G., Mateus, A., \& Vale Asari, N.\ 2011, \mnras, 413, 1687 
\bibitem[Combes(2017)]{Combes2017} Combes, F.\ 2017, Frontiers in Astronomy and Space Sciences, 4, 10 
\bibitem[Croton et al.(2006)]{Croton2006} Croton, D.~J., Springel, V., White, S.~D.~M., et al.\ 2006, \mnras, 365, 11 
\bibitem[Croxall et al.(2012)]{Croxall2012} Croxall, K.~V., Smith, J.~D., Wolfire, M.~G., et al.\ 2012, \apj, 747, 81 
\bibitem[da Cunha et al.(2008)]{da Cunha2008} da Cunha, E., Charlot, S., \& Elbaz, D.\ 2008, \mnras, 388, 1595 
\bibitem[Dale et al.(2007)]{Dale2007} Dale, D.~A., Gil de Paz, A., Gordon, K.~D., et al.\ 2007, \apj, 655, 863 
\bibitem[Davis et al.(2013)]{Davis2013} Davis, T.~A., Alatalo, K., Bureau, M., et al.\ 2013, \mnras, 429, 534 
\bibitem[Davis et al.(2014)]{Davis2014} Davis, T.~A., Young, L.~M., Crocker, A.~F., et al.\ 2014, \mnras, 444, 3427 
\bibitem[de Vaucouleurs et al.(1991)]{deVauc91} de Vaucouleurs, 
G., de Vaucouleurs, A., Corwin, H.~G., Jr., et al.\ 1991, Third Reference 
Catalogue of Bright Galaxies.~Vol. I,~ Vol. II,~ Vol III (New York: Springer), 2091
\bibitem[Dekel \& Birnboim(2006)]{Dekel2006} Dekel, A., \& Birnboim, Y.\ 2006, \mnras, 368, 2 
\bibitem[Di Matteo et al.(2005)]{Di2005} Di Matteo, T., Springel, V., \& Hernquist, L.\ 2005, \nat, 433, 604 
\bibitem[D{\'{\i}}az-Santos et al.(2013)]{Diaz-Santos2013} D{\'{\i}}az-Santos, T., Armus, L., Charmandaris, V., et al.\ 2013, \apj, 774, 68 
\bibitem[D{\'{\i}}az-Santos et al.(2017)]{Diaz-Santos2017} D{\'{\i}}az-Santos, T., Armus, L., Charmandaris, V., et al.\ 2017, arXiv:1705.04326 
\bibitem[Draine et al.(2007)]{Draine2007} Draine, B.~T., Dale, D.~A., Bendo, G., et al.\ 2007, \apj, 663, 866 
\bibitem[Elbaz et al.(2007)]{Elbaz2007} Elbaz, D., Daddi, E., Le Borgne, D., et al.\ 2007, \aap, 468, 33 
\bibitem[Fang et al.(2013)]{Fang2013} Fang, J.~J., Faber, S.~M., Koo, D.~C., \& Dekel, A.\ 2013, \apj, 776, 63 
\bibitem[Farrah et al.(2013)]{Farrah2013} Farrah, D., Lebouteiller, V., Spoon, H.~W.~W., et al.\ 2013, \apj, 776, 38 
\bibitem[Galametz et al.(2013)]{Galametz2013} Galametz, M., Kennicutt, R.~C., Calzetti, D., et al.\ 2013, \mnras, 431, 1956 
\bibitem[Goldsmith et al.(2015)]{Goldsmith2015} Goldsmith, P.~F., Y{\i}ld{\i}z, U.~A., Langer, W.~D., \& Pineda, J.~L.\ 2015, \apj, 814, 133 
\bibitem[Gordon et al.(2008)]{Gordon2008} Gordon, K.~D., Engelbracht, C.~W., Rieke, G.~H., et al.\ 2008, \apj, 682, 336-354 
\bibitem[Griffin et al.(2010)]{Griffin2010} Griffin, M.~J., Abergel, A., Abreu, A., et al.\ 2010, \aap, 518, L3 
\bibitem[Gunn et al.(1998)]{Gunn1998} Gunn, J.~E., Carr, M., 
Rockosi, C., et al.\ 1998, \aj, 116, 3040 
\bibitem[Habing(1968)]{Habing1968} Habing, H.~J.\ 1968, \bain, 19, 421 
\bibitem[Ho et al.(1997)]{Ho1997} Ho, L.~C., Filippenko, A.~V., \& Sargent, W.~L.~W.\ 1997, \apjs, 112, 315 
\bibitem[Hopkins et al.(2006)]{Hopkins2006} Hopkins, P.~F., Hernquist, L., Cox, T.~J., et al.\ 2006, \apjs, 163, 1 
\bibitem[Hughes et al.(2015)]{Hughes2015} Hughes, T.~M., Foyle, K., Schirm, M.~R.~P., et al.\ 2015, \aap, 575, A17 
\bibitem[Jarrett et al.(2000)]{Jarrett2000} Jarrett, T.~H., Chester, T., Cutri, R., et al.\ 2000, \aj, 119, 2498 
\bibitem[Kamenetzky et al.(2016)]{Kamenetzky2016} Kamenetzky, J., Rangwala, N., Glenn, J., Maloney, P.~R., \& Conley, A.\ 2016, \apj, 829, 93 
\bibitem[Kaufman et al.(1999)]{Kaufman1999} Kaufman, M.~J., Wolfire, M.~G., Hollenbach, D.~J., \& Luhman, M.~L.\ 1999, \apj, 527, 795 
\bibitem[Kauffmann et al.(2003)]{Kauffmann2003} Kauffmann, G., 
Heckman, T.~M., Tremonti, C., et al.\ 2003, \mnras, 346, 1055 
\bibitem[Kaufman et al.(2006)]{Kaufman2006} Kaufman, M.~J., Wolfire, M.~G., \& Hollenbach, D.~J.\ 2006, \apj, 644, 283 
\bibitem[Kelz et al.(2006)]{Kelz2006} Kelz, A., Verheijen, M.~A.~W., Roth, M.~M., et al.\ 2006, \pasp, 118, 129 
\bibitem[Kennicutt(1998a)]{Kennicutt1998a} Kennicutt, R.~C., Jr.\ 1998a, \apj, 498, 541 
\bibitem[Kennicutt(1998b)]{Kennicutt1998b} Kennicutt, R.~C., Jr.\ 1998b, \araa, 36, 189 
\bibitem[Kere{\v s} et al.(2005)]{keres2005} Kere{\v s}, D., Katz, N., Weinberg, D.~H., \& Dav{\'e}, R.\ 2005, \mnras, 363, 2 
\bibitem[Kewley et al.(2001)]{Kewley2001} Kewley, L.~J., Dopita, M.~A., Sutherland, R.~S., Heisler, C.~A., \& Trevena, J.\ 2001, \apj, 556, 121 
\bibitem[Kewley \& Dopita(2002)]{Kewley2002} Kewley, L.~J., \& Dopita, M.~A.\ 2002, \apjs, 142, 35 
\bibitem[Krajnovi{\'c} et al.(2011)]{Krajnovic2011} Krajnovi{\'c}, D., Emsellem, E., Cappellari, M., et al.\ 2011, \mnras, 414, 2923 
\bibitem[Kramer et al.(2005)]{Kramer2005} Kramer, C., Mookerjea, B., Bayet, E., et al.\ 2005, \aap, 441, 961 
\bibitem[Lanz et al.(2016)]{Lanz2016} Lanz, L., Ogle, P.~M., Alatalo, K., \& Appleton, P.~N.\ 2016, \apj, 826, 29 
\bibitem[Lapham et al.(2017)]{Lapham2017} Lapham, R.~C., Young, L.~M., \& Crocker, A.\ 2017, \apj, 840, 51 
\bibitem[Leroy et al.(2008)]{Leroy2008} Leroy, A.~K., Walter, F., Brinks, E., et al.\ 2008, \aj, 136, 2782 
\bibitem[Leroy et al.(2013)]{Leroy2013} Leroy, A.~K., Walter, F., Sandstrom, K., et al.\ 2013, \aj, 146, 19 
\bibitem[Li \& Draine(2001)]{Li2001} Li, A., \& Draine, B.~T.\ 2001, \apj, 554, 778 
\bibitem[Loubser \& S{\'a}nchez-Bl{\'a}zquez(2011)]{Loubser2011} Loubser, S.~I., \& S{\'a}nchez-Bl{\'a}zquez, P.\ 2011, \mnras, 410, 2679 
\bibitem[Lu et al.(2017)]{Lu2017} Lu, N., Zhao, Y., D{\'{\i}}az-Santos, T., et al.\ 2017, \apjs, 230, 1 
\bibitem[Malhotra et al.(2000)]{Malhotra2000} Malhotra, S., Hollenbach, D., Helou, G., et al.\ 2000, \apj, 543, 634 
\bibitem[Malhotra et al.(2001)]{Malhotra2001} Malhotra, S., Kaufman, M.~J., Hollenbach, D., et al.\ 2001, \apj, 561, 766 
\bibitem[Martig et al.(2009)]{Martig2009} Martig, M., Bournaud, F., Teyssier, R., \& Dekel, A.\ 2009, \apj, 707, 250 
\bibitem[McDermid et al.(2015)]{McDermid2015} McDermid, R.~M., Alatalo, K., Blitz, L., et al.\ 2015, \mnras, 448, 3484 
\bibitem[Moshir et al.(1990)]{Moshir1990} Moshir, M., et al.\ 1990, IRAS Faint Source Catalogue, version 2.0 (1990) 
\bibitem[Negishi et al.(2001)]{Negishi2001} Negishi, T., Onaka, T., Chan, K.-W., \& Roellig, T.~L.\ 2001, \aap, 375, 566 
\bibitem[Nesvadba et al.(2010)]{Nesvadba2010} Nesvadba, N.~P.~H., Boulanger, F., Salom{\'e}, P., et al.\ 2010, \aap, 521, A65 
\bibitem[Noll et al.(2009)]{Noll2009} Noll, S., Burgarella, D., Giovannoli, E., et al.\ 2009, \aap, 507, 1793 
\bibitem[Nyland et al.(2016)]{Nyland2016} Nyland, K., Young, L.~M., Wrobel, J.~M., et al.\ 2016, \mnras, 458, 2221 
\bibitem[Oberst et al.(2006)]{Oberst2006} Oberst, T.~E., Parshley, S.~C., Stacey, G.~J., et al.\ 2006, \apjl, 652, L125 
\bibitem[Oberst et al.(2011)]{Oberst2011} Oberst, T.~E., Parshley, S.~C., Nikola, T., et al.\ 2011, \apj, 739, 100 
\bibitem[Ogle et al.(2007)]{Ogle2007} Ogle, P., Antonucci, R., Appleton, P.~N., \& Whysong, D.\ 2007, \apj, 668, 699 
\bibitem[Onishi et al.(2017)]{Onishi2017} Onishi, K., Iguchi, S., Davis, T.~A., et al.\ 2017, arXiv:1703.05247 
\bibitem[Osterbrock(1989)]{Osterbrock1989} Osterbrock, D.~E.\ 1989, Astronomy, 17, 102 
\bibitem[Ott (2010)]{Ott2010} Ott, S.\ 2010, Astronomical Data Analysis Software and Systems XIX, 434, 139 
\bibitem[Parkin et al.(2013)]{Parkin2013} Parkin, T.~J., Wilson, C.~D., Schirm, M.~R.~P., et al.\ 2013, \apj, 776, 65 
\bibitem[Parkin et al.(2014)]{Parkin2014} Parkin, T.~J., Wilson, C.~D., Schirm, M.~R.~P., et al.\ 2014, \apj, 787, 16
\bibitem[Peng et al.(2002)]{Peng2002} Peng, C.~Y., Ho, L.~C., Impey, C.~D., \& Rix, H.-W.\ 2002, \aj, 124, 266  
\bibitem[Peng et al.(2010)]{Peng2010} Peng, C.~Y., Ho, L.~C., Impey, C.~D., \& Rix, H.-W.\ 2010, \aj, 139, 2097 
\bibitem[Pilbratt et al.(2010)]{pil10}Pilbratt, G.L., Riedinger, J. R., Passvogel, T. et al. 2010, A\&A, 518, L1
\bibitem[Poglitsch et al.(2010)]{Poglitsch2010} Poglitsch, A., Waelkens, C., Geis, N., et al.\ 2010, \aap, 518, L2 
\bibitem[Pound \& Wolfire(2008)]{Pound2008} Pound, M.~W., \& Wolfire, M.~G.\ 2008, Astronomical Data Analysis Software and Systems XVII, 394, 654 
\bibitem[Roth et al.(2005)]{Roth2005} Roth, M.~M., Kelz, A., Fechner, T., et al.\ 2005, \pasp, 117, 620 
\bibitem[Sanders \& Mirabel(1996)]{Sanders1996} Sanders, D.~B., \& Mirabel, I.~F.\ 1996, \araa, 34, 749 
\bibitem[Sargsyan et al.(2014)]{Sargsyan2014} Sargsyan, L., Samsonyan, A., Lebouteiller, V., et al.\ 2014, \apj, 790, 15 
\bibitem[Savage \& Sembach(1996)]{Savage1996} Savage, B.~D., \& Sembach, K.~R.\ 1996, \araa, 34, 279 
\bibitem[Serra et al.(2012)]{Serra2012} Serra, P., Oosterloo, T., Morganti, R., et al.\ 2012, \mnras, 422, 1835 
\bibitem[Shi et al.(2011)]{Shi2011} Shi, Y., Helou, G., Yan, L., et al.\ 2011, \apj, 733, 87 
\bibitem[Shi et al.(2014)]{Shi2014} Shi, Y., Armus, L., Helou, G., et al.\ 2014, \nat, 514, 335 
\bibitem[Shi et al.(2018)]{Shi2018} Shi, Y., Yan, L., Armus, L., et al.\ 2018, arXiv:1801.00888
\bibitem[Smethurst et al.(2016)]{Smethurst2016} Smethurst, R.~J., Lintott, C.~J., Simmons, B.~D., et al.\ 2016, \mnras, 463, 2986 
\bibitem[Smith et al.(2012)]{Smith2012} Smith, M.~W.~L., Gomez, H.~L., Eales, S.~A., et al.\ 2012, \apj, 748, 123
\bibitem[Stacey et al.(1991)]{Stacey1991} Stacey, G.~J., Geis, N., Genzel, R., et al.\ 1991, \apj, 373, 423 
\bibitem[Tacchella et al.(2015)]{Tacchella2015} Tacchella, S., Carollo, C.~M., Renzini, A., et al.\ 2015, Science, 348, 314 
\bibitem[Temi et al.(2009a)]{Temi2009a} Temi, P., Brighenti, F., \& Mathews, W.~G.\ 2009, \apj, 695, 1 
\bibitem[Temi et al.(2009b)]{Temi2009b} Temi, P., Brighenti, F., \& Mathews, W.~G.\ 2009, \apj, 707, 890 
\bibitem[Temi et al.(2007a)]{Temi2007a} Temi, P., Brighenti, F., \& Mathews, W.~G.\ 2007a, \apj, 660, 1215 
\bibitem[Temi et al.(2007b)]{Temi2007b} Temi, P., Brighenti, F., \& Mathews, W.~G.\ 2007b, \apj, 666, 222 
\bibitem[Tielens \& Hollenbach(1985)]{Tielens1985} Tielens, A.~G.~G.~M., \& Hollenbach, D.\ 1985, \apj, 291, 722 
\bibitem[Toomre(1964)]{Toomre1964} Toomre, A.\ 1964, \apj, 139, 1217 
\bibitem[Tremonti et al.(2004)]{Tremonti2004} Tremonti, C.~A., Heckman, T.~M., Kauffmann, G., et al.\ 2004, \apj, 613, 898 
\bibitem[Veilleux \& Osterbrock(1987)]{Veilleux1987} Veilleux, S., \& Osterbrock, D.~E.\ 1987, \apjs, 63, 295
\bibitem[Verheijen et al.(2004)]{Verheijen2004} Verheijen, M.~A.~W., Bershady, M.~A., Andersen, D.~R., et al.\ 2004, Astronomische Nachrichten, 325, 151 
\bibitem[Wolfire et al.(1990)]{Wolfire1990} Wolfire, M.~G., Tielens, A.~G.~G.~M., \& Hollenbach, D.\ 1990, \apj, 358, 116 
\bibitem[Xiao et al.(2016)]{Xiao2016} Xiao, M.-Y., Gu, Q.-S., Chen, Y.-M., \& Zhou, L.\ 2016, \apj, 831, 63 
\bibitem[Young et al.(2011)]{Young2011} Young, L.~M., Bureau, M., Davis, T.~A., et al.\ 2011, \mnras, 414, 940 
\bibitem[Young et al.(2014)]{Young2014} Young, L.~M., Scott, N., Serra, P., et al.\ 2014, \mnras, 444, 3408 
\bibitem[Zhao et al.(2013)]{Zhao2013} Zhao, Y., Lu, N., Xu, C.~K., et al.\ 2013, \apjl, 765, L13 
\bibitem[Zhao et al.(2016a)]{Zhao2016a} Zhao, Y., Lu, N., Xu, C.~K., et al.\ 2016a, \apj, 819, 69 
\bibitem[Zhao et al.(2016b)]{Zhao2016b} Zhao, Y., Yan, L., \& Tsai, C.-W.\ 2016b, \apj, 824, 146 
\bibitem[Zubko et al.(2004)]{Zubko2004} Zubko, V., Dwek, E., \& Arendt, R.~G.\ 2004, \apjs, 152, 211 






\end{thebibliography}
\end{document}